\documentclass[a4paper,10pt]{extarticle}
\usepackage{extsizes}
\usepackage[utf8]{inputenc}
\usepackage{amsmath,amssymb}
\usepackage{authblk}
\usepackage{float}
\usepackage{array}
\usepackage{mathrsfs}
\usepackage{setspace}
\usepackage{subfig}
\usepackage{cancel}
\usepackage{graphicx}
\usepackage{txfonts}
\usepackage{graphicx}
\usepackage{anysize}
\usepackage{hyperref}
\usepackage{textcomp}
\usepackage{caption}
\usepackage{float}
\usepackage{epsfig}
\usepackage[normalem]{ulem}
\title{Strategies to cure numerical shock instability in HLLEM Riemann solver}
\author[1]{Sangeeth Simon}
\author[2]{J. C. Mandal \thanks{Corresponding author: mandal@iitb.ac.in}}

\affil[1,2]{Department of Aerospace Engineering, Indian Institute of Technology Bombay, Mumbai-400076}
\date{}
\begin{document}

\maketitle
%
%

\begin{abstract}
The HLLEM scheme is a popular contact and shear preserving approximate Riemann solver for cheap and accurate computation of high speed 
gasdynamical flows.
Unfortunately this scheme is known to be plagued by various forms of numerical shock instability. In this paper we present various strategies to save the HLLEM scheme from developing such spurious solutions.
A linear scale analysis of its mass and interface-normal momentum flux discretizations reveal that its antidiffusive terms, which are primarily responsible for 
resolution of linear wavefields, are inadvertently activated along a normal shock front due to numerical perturbations.
These erroneously activated terms counteract the favourable damping mechanism provided by its
inherent HLL-type diffusive terms and trigger the shock instability.
To avoid this, two different strategies are proposed for discretization of these critical flux components in the vicinity of a shock: one that deals with 
increasing the magnitude of inherent HLL-type dissipation through careful manipulation of specific non-linear wave speed estimates while the other deals
with reducing the magnitude of these critical antidiffusive terms.   
A linear perturbation analysis is performed to 
gauge the effectiveness of these cures and estimate von-Neumann type stability bounds on the CFL number arising from their use.
Results from classic numerical test cases show that both types of modified HLLEM schemes are able to provide excellent shock stable solutions while 
retaining commendable accuracy on shear dominated viscous flows.
\end{abstract}

\section{Introduction}

Computation of high speed gasdynamical flows has benefited tremendously from the development of various approximate Riemann solvers for the Euler system of equations. Although most of these schemes
are designed to capture nonlinear waves like shocks and expansion fans, they can be broadly classifed into two groups based on their ability to capture the linear waves like  entropy and shear waves 
accurately. The resulting two classes of approximate Riemann solvers are generally referred to in literature as contact-shear preserving and contact-shear dissipative solvers respectively. 
Roe scheme \cite{roe1981} is one of the most popular examples of the former category and is widely incorporated in most industry and research codes. However, Roe scheme has several problems
like requirement of an entropy fix to resolve sonic expansions, lack of positivity in density and internal energy in near vaccum flows and requirement of the knowledge of
full eigen structure of the flux jacobians. On the other hand a popular example of a contact-shear dissipative type is the HLL scheme \cite{hll1983,einfeldt1988}. Since these are based 
directly on integral form of conservation laws, they do not require linearizations or complete knowledge of eigenstructures of the Euler system. Consequently, these 
methods do not suffer
from the issues mentioned here that plagues the Roe scheme. However lack of contact and shear ability seriously restricts their applicability to practical problems involving shearing flows, boundary layer
flows, flame and material fronts etc \cite{toro_vaz2012,vanleer1987}. Einfeldt et al \cite{einfeldt1991} proposed an interesting option that combines the advantages 
of the Roe scheme with that of the HLL scheme called HLLEM scheme. The HLLEM scheme is positivity preserving, contact-shear capturing and does not require an entropy fix. 
Also it demands knowledge of only the eigenstructure of the linear wavefields of the system. 

Unfortunately like other contact-shear preserving Riemann solvers, the HLLEM scheme also suffers from various forms of numerical shock instability during 
simulations of multidimensional flows which involves regions of strong grid aligned normal shocks. A classic example of such a failure is the 
Carbuncle phenomenon \cite{peery1988}. 
A catalogue of such failures can be found in \cite{quirk1994}. Benign perturbations in the initial conditions are rapidly amplified in the numerical shock region leading 
to spurious solutions. Most commonly, these instabilities manifest as oscillations in the shock profile, polluted after shock values, growth in error norms in case of 
steady state problems and in extreme cases complete breakdown of the solution. 

Although there is still no agreement on the exact cause of these spurious solutions, many authors suggest that lack of cross-flow dissipation provided by the contact-shear 
preserving 
Riemann solvers along the shock front could be a major trigger \cite{sanders1998,shen2014}. A loss in overall dissipation for a contact-shear 
preserving scheme in such circumstances are 
argued to happen due to vanishing of their numerical dissipation terms that corresponds to accuracy on linear waves \cite{shen2014,ren2003}. On the other hand, a linear wave dissipative scheme
like the HLLE does not suffer from this.
Many cures for 
creating a shock stable Roe scheme, for example, are hence based on 
increasing the component of dissipation corresponding to the linear waves in the overall dissipation matrix \cite{sanders1998,ren2003,lin1995,pandolfi2001,kim2003,farhad2006,nishikawa2008}.
Similarly, in case of the HLLEM scheme, Park et al \cite{park2003} proposed a control of the antidiffusive terms, which are chiefly responsible for
restoring accuracy on linear wavefields, to ensure shock stability. Xie et al \cite{xie2017} showed that merely controlling the antidiffusive terms corresponding to 
the shear wave could guarantee shock stability. Obayashi et al \cite{obayashi1994} proposed a Carbuncle free HLLEM scheme based on retaining only its HLLE component in 
the vicinity of a shock although their scheme could benefit from more rigorous testing.

Recently it was clarified in case of another popular HLL-based contact-shear preserving Riemann solver called HLLC scheme, that the antidiffusive terms, specifically in mass and 
interface-normal momentum flux
discretizations in transverse direction of a shock front are the major triggers for instability \cite{sangeeth2018_HLLCADC}. It has been found that the antidiffusive terms present in the HLLC 
discretization of these critical flux components gets activated in the transverse direction of a numerically perturbed normal shock front and causes a 
reduction in the magnitude of 
its inherent HLL-type diffusive terms. The weakened diffusive terms fail to provide requisite damping of perturbations in $\rho$ and $u$ variables resulting in 
unphysical variation of conserved quantity $\rho u$ along the shock front and subsequent development of shock unstable solutions.
Based on this, the authors suggested an inexpensive shock stable HLLC scheme 
by simply controlling the antidiffusive terms in these critical flux components using a simple differentiable pressure sensor. 
An interesting alternative to save the HLLC scheme that does not rely upon controlling the antidiffusive terms was proposed in \cite{sangeeth2018_HLLCSLSR}. 
It involves a strategy to increase the inherent HLL-type 
dissipation of this scheme in the vicinity of shock through careful modification of certain nonlinear wavespeed estimates. Since the HLLC scheme and the HLLEM scheme 
are both derived from the HLL scheme, they share identical HLL-type diffusive terms. Thus, it opens up the possibility of extending these appealing cures developed for the former 
to the latter provided they share similar instability-trigger mechanism.



In this paper, we first perform a linear scale analysis of the
numerical dissipation terms of the HLLEM scheme appearing in the mass and interface-normal momentum flux discretizations in the vicinity of a steady shock front subjected 
to numerical perturbations with an objective to identify specific terms in its formulation that triggers the instability.  
Our analysis indicates that the instability in the HLLEM scheme is also promoted due to a weakening of 
its embedded HLL-type dissipation in the direction parallel to a shock front due to erroneous activation of its antidiffusive terms. 
Based on these findings, we propose two robust strategies to save the HLLEM scheme from shock instability that are based on methods suggested in \cite{sangeeth2018_HLLCADC,sangeeth2018_HLLCSLSR}.
The first method focuses on increasing the inherent HLL-type numerical dissipation existing within the HLLEM scheme. 
This is achieved by carefully modifying non linear wave speed estimates of this embedded HLLE dissipation. Two multidimensional local solution dependent 
strategies are discussed to ascertain the quantity of supplementary dissipation to be provided to enhance this HLLE component.
The second method focuses on reducing the magnitude of the antidiffusive component of the HLLEM scheme relative to its inherent HLL-type dissipation.
A technique of linear perturbation analysis 
\cite{pandolfi2001} is used to show how these different strategies are able to easily dampen numerical perturbations that may appear in the primitive quantities.
Corresponding von Neumann stability bounds on the allowable CFL number for implementation of all of these strategies are also estimated.
We also demonstrate
through our analysis and numerical examples how simply controlling the antidiffusive terms corresponding to shear waves alone, as suggested in \cite{xie2017},
may not suffice to create a shock stable 
HLLEM scheme.  Finally, a suite of first order and 
second order test cases are used to study the effectiveness of these methods in dealing with shock instabilities.

This paper is organized as follows. In Sec.(\ref{sec:governingequations},\ref{sec:FVM_framework}) we brief the governing equations and finite volume framework 
adopted in this paper. Sec.(\ref{sec:recapofhlleandhllem}) recapitulates the details of the HLLE and the HLLEM schemes that form the basis of our discussion. In 
Sec.(\ref{sec:orderofmagnitudeanalysis_hlle_hllem}) we use a linear scale analysis on the HLLE and HLLEM schemes to identify the specific terms in the HLLEM scheme that 
triggers the instability.
This analysis is supplemented with a linear perturbation analysis that provides insights on the evolution of perturbations of primitive quantities in these schemes.
Based on these, we describe strategies to formulate a few shock stable versions of the HLLEM scheme in Sec.(\ref{sec:formulations}). The linear perturbation analysis developed
earlier is extended to study the behavior of these new shock stable versions in dealing with the perturbations. In Sec.(\ref{sec:numericaltests}) certain classic shock 
instability test cases are used to demonstrate the robustness of these new formulations. Sec(\ref{sec:conclusions}) presents some concluding remarks.

\section{Governing equations}
\label{sec:governingequations}
The governing equations for two dimensional inviscid compressible flow can be expressed in their conservative form as, 
\begin{align}
 \frac{\partial \mathbf{\acute{U}}}{\partial t} + \frac{\partial \mathbf{\acute{F}(U)}}{\partial x} + \frac{\partial \mathbf{\acute{G}(U)}}{\partial y} = 0
 \label{equ:EE-differentialform}
\end{align}
where $\mathbf{\acute{U}}, \mathbf{\acute{F}(U)} , \mathbf{\acute{G}(U)}$ are the vector of conserved variables and x and y directional fluxes respectively. These are 
given by, 

\begin{align}
 \mathbf{\acute{U}} = \left [ \begin{array}{c}
\rho \\
\rho u\\
\rho v\\
\rho E
\end{array}\right ],
\mathbf{\acute{F}(U)} = \left [ \begin{array}{c}
\rho u \\
\rho u^2 + p\\
\rho u v\\
(\rho E + p)u
\end{array}\right ],
\mathbf{\acute{G}(U)}= \left [ \begin{array}{c}
\rho v \\
\rho u v\\
\rho v^2 + p\\
(\rho E + p)v
\end{array}\right ]
\end{align}
In the above expressions, $\rho, u, v, p$ and $E$ stands for density, x-velocity, y-velocity, pressure and specific total energy. The system of 
equations are closed through the equation of state,
\begin{align}
 p = (\gamma-1)\left(\rho E - \frac{1}{2}\rho(u^2 + v^2)\right)
\end{align}
where $\gamma$ is the ratio of specific heats. Present work assumes a calorifically perfect gas with $\gamma =1.4$. A particularly useful form of the 
Eq.(\ref{equ:EE-differentialform}) is the integral form given by,

\begin{align}
 \frac{\partial}{\partial t} \int_{\varOmega} \mathbf{\acute{U}} dxdy + \oint_{d\varOmega} [(\mathbf{\acute{F},\acute{G}}).\mathbf{n}] dl = 0 
 \label{eqn:EE-integralform}
\end{align}
where $\varOmega$ denotes a control volume over which Eq.(\ref{eqn:EE-integralform}) describes a Finite Volume balance of the conserved
quantities, $dx$ and $dy$
denotes the x and y dimensions of the control volume respectively, $d\varOmega$ denotes the boundary 
surface of the control volume and $dl$ denotes an infinitesimally small element on $d\varOmega$. $\mathbf{n}$ is the outward pointing unit normal vector to the surface $d\varOmega$. 

\section{Finite volume discretization}
\label{sec:FVM_framework}
\begin{figure}[H]Hence i
	    \centering
	    \includegraphics[scale=0.3]{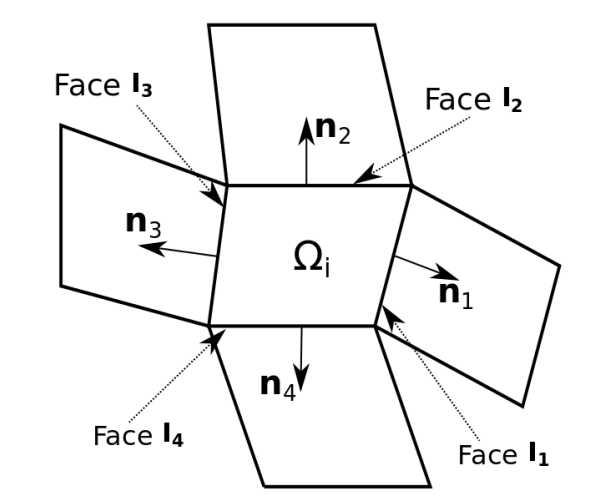}
	    \caption{Typical control volume $i$ with its associated interfaces $I_k$ and respective normal vectors $\mathbf{n}_k$.}
	    \label{fig:finitevolume}
	    \end{figure}

In this paper we seek a Finite Volume numerical solution of Eq.(\ref{eqn:EE-integralform}) by discretizing the equation on a computational 
mesh consisting of structured quadrilateral cells as shown in Fig.(\ref{fig:finitevolume}). For a typical cell $i$ belonging to this mesh, a 
semi-discretized version
of the governing equation can be written as,

\begin{align}
 \frac{d\mathbf{U}_i}{dt} = -\frac{1}{|\varOmega|_i} \sum_{k=1}^{4} [({\mathbf{\acute{F},\acute{G}}})_k.\mathbf{n}_k] \Delta s_k
 \label{eqn:EE-partialdiscretized}
\end{align}
where ${\mathbf{U}_i}$ is an appropriate cell averaged conserved state vector, $({\mathbf{\acute{F},\acute{G}}})_k$ denotes the flux vector at the 
mid point of each interface $I_k$ while $\mathbf{n}_k$ and $\Delta s_k$ denotes the unit normal vector  and the length of each $I_k$ interface 
respectively. These are shown in Fig.(\ref{fig:finitevolume}). The interface flux $({\mathbf{\acute{F},\acute{G}}})_k.\mathbf{n}_k$ can be obtained by various methods. 
One of the most popular class of methods 
to compute this are the approximate Riemann solvers. A conventional two state approximate Riemann solver uses the 
rotational invariance property of Euler equations to express the term $({\mathbf{\acute{F},\acute{G}}})_k.\mathbf{n}_k$ as, 

\begin{align}
 \frac{d\mathbf{U}_i}{dt} = -\frac{1}{|\varOmega|_i} \sum_{k,m=1}^{4} [\mathbf{T}^{-1}_{k} \mathbf{F} (\mathbf{U_L},\mathbf{U_R})]\Delta s_k
\label{eqn:EE-rotationalinvariance}
 \end{align}
 where $ \mathbf{U_L} = \mathbf{T}_{k}(\mathbf{U}_i), \mathbf{U_R}=\mathbf{T}_{k}(\mathbf{U}_m)$ indicates the initial conditions of a local
Riemann problem across $k^{th}$ interface shared by the cells $i$ and $m$. The matrices $\mathbf{T}_{k}$ and $\mathbf{T}_{k}^{-1}$ are rotation matrices at the $k^{th}$ interface given by,
 

\begin{align}
 \mathbf{T}_k= \left [ \begin{array}{r}
1\\
0\\
0\\
0
\end{array}
\begin{array}{r}
0\\
n_{xk}\\
-n_{yk}\\
0
\end{array}
\begin{array}{r}
0\\
n_{yk}\\
n_{xk}\\
0 
\end{array}
\begin{array}{r}
0\\
0\\
0\\
1 
\end{array}\right ],\mathbf{T}^{-1}_k= \left [ \begin{array}{r}
1\\
0\\
0\\
0
\end{array}
\begin{array}{r}
0\\
n_{xk}\\
n_{yk}\\
0
\end{array}
\begin{array}{r}
0\\
-n_{yk}\\
n_{xk}\\
0 
\end{array}
\begin{array}{r}
0\\
0\\
0\\
1 
\end{array}\right ] 
\end{align}
where $n_{xk},n_{yk}$ denote the components of the normal vector $\mathbf{n}$.

\noindent In the next section we briefly describe two approximate Riemann solvers named the HLLE scheme and the HLLEM scheme that can be used to estimate the flux 
$\mathbf{F} (\mathbf{U_L},\mathbf{U_R})$
at a given interface. To avoid extra notations, henceforth $\mathbf{F}$ simply represent the local Riemann flux at any interface with outward pointing normal $\mathbf{n}$. 
\section{Recap of HLLE and HLLEM schemes}
\label{sec:recapofhlleandhllem}
The original HLL scheme was devised by Harten, Lax and van Leer \cite{hll1983}. Unlike linearized solver of Roe, the HLL scheme does not require 
the knowledge of the eigenstructure of the Euler system. Instead it assumes a wave structure consisting of two waves that seperates three constant states.
Using the integral form of the conservation laws on this wave structure, the HLL Riemann flux can then be written as, 

\begin{align}
    \mathbf{F}_{HLL}= \frac{S_R \mathbf{F}_L - S_L \mathbf{F}_R + S_LS_R (\mathbf{U}_R - \mathbf{U}_L) }{S_R - S_L}
\label{eqn:hllformualtion}
\end{align}
Here $\mathbf{F}_L = \mathbf{F(\mathbf{U}_L)}$
and $\mathbf{F}_R = \mathbf{F(\mathbf{U}_R)}$ are the local fluxes at either side of the interface.
$S_L$ and $S_R$ are numerical approximations to the speeds of the left most and right most running characteristics that emerge as the solution of 
the Riemann problem at an interface.
It has been shown that under appropriate choice of wavespeeds $S_L$ and $S_R$, the HLL scheme is both positivity preserving and entropy
satisfying \cite{einfeldt1988}. This choice of wavespeeds are given as, 

\begin{align}
\nonumber
 S_L = min(0,u_{nL}-a_L, \tilde{u}_n-\tilde{a})\\
 S_R = max(0,u_{nR}+a_R, \tilde{u}_n +\tilde{a})
 \label{eqn:HLLEwavespeedestimate}
\end{align}
where $u_{nL,R}$ are the normal velocity across an interface, $a_{L,R}$ are the respective sonic speeds and $\tilde{u}_n,\tilde{a}$ are the standard Roe averaged 
quantities at the interface \cite{roe1981}. Using these wavespeed estimates, the HLL scheme is also known as the 
HLLE scheme \cite{einfeldt1988}. For the purpose of this paper, we note that HLLE scheme can also be rewritten as, 
\begin{align}
   \mathbf{F}_{HLLE} = \frac{1}{2}(\mathbf{F}_L + \mathbf{F}_R) + \mathbf{D}
   \label{eqn:HLLE_centralplusdissipation}
\end{align}
where the dissipation can be written as \cite{mandal2012},
\begin{align}
 \mathbf{D} = \frac{S_R+S_L}{2(S_R-S_L)}(\mathbf{F}_L-\mathbf{F}_R) + \frac{S_LS_R}{(S_R-S_L)}(\mathbf{U}_R-\mathbf{U}_L)
 \label{eqn:HLLEdissipation}
\end{align}
Although quite accurate in resolution of shocks and expansion fans, the major drawback of HLL scheme is its inability to 
resolve the contact and shear waves. The loss of accuracy on these waves occur because of the assumption of constant average state between the two waves.
Further, it is observed that HLL Riemann solver is free from various forms of numerical shock instability which is generally attributed to its 
linear wave dissipative nature \cite{pandolfi2001, gressier2000}.

\noindent Einfeldt et al \cite{einfeldt1991} presented an improved version of the HLLE scheme that can capture the contact and shear waves in addition to retaining the excellent positivity
preserving and entropy satisfying property of the original HLLE scheme. Additional accuracy on these waves for HLLE scheme were obtained by explicitly adding necessary 
antidiffusive terms corresponding to only linear waves. The HLLEM flux can be written as, 
\begin{align}
\mathbf{F}_{HLLEM}= \frac{S_R \mathbf{F}_L - S_L \mathbf{F}_R + S_LS_R (\mathbf{U}_R - \mathbf{U}_L) } {S_R - S_L} + \frac{S_LS_R}{S_R-S_L}(- \delta_{2} \tilde{\alpha}_{2} \tilde{\mathbf{R}}_{2} - \delta_{3} 
\tilde{\alpha}_{3} \tilde{\mathbf{R}}_{3})
\label{eqn:hllemformualtion}
\end{align}
with wave speeds $S_L,S_R$ still defined by Eq.(\ref{eqn:HLLEwavespeedestimate}). Here, terms $ \tilde{\mathbf{R}}_{2,3},\tilde{\alpha}_{2,3}$ denote respectively the 
right eigenvectors of the flux jacobians corresponding to the linearly degenerate waves and the strength of these waves obtained
by projecting the total jump $\mathbf{U}_R-\mathbf{U}_L$ onto these eigenvectors. Both these quantities are evaluated at the Roe averaged state. These are given as \cite{toro2009}, 

\begin{align}
\tilde{\mathbf{R}}_{2}= \left [ \begin{array}{c}
1 \\
\tilde{u}_n\\
\tilde{u}_t\\
\frac{1}{2}(\tilde{u}_n^2+\tilde{u}_t^2)
\end{array}\right ],\ \ \tilde{\mathbf{R}}_{3}= \left [ \begin{array}{c}
0 \\
0\\
1\\
\tilde{u}_t
\end{array}\right ] \ \ 
\label{eqn:HLLEM_lineareigenvectors}
\end{align}

\begin{align}
 \tilde{\alpha}_{2} = \Delta \rho - \frac{\Delta p}{\tilde{a}^2},\ \  \tilde{\alpha}_{3} = -\tilde{\rho} \Delta {u}_t
 \label{eqn:HLLEM_linearwavestrengths}
\end{align}
where $\Delta (.)$ operator represents the operation $(.)_R - (.)_L$, $u_t$ denotes the tangential component of velocity at an interface and $\tilde{u}_t$ denotes the corresponding
Roe averaged quantity.
The terms $\delta_{2,3}$ represent coefficients that control the amount of antidissipation being introduced into these waves. An estimate for these that guarantees 
good accuracy for resolving linear waves are given as \cite{park2003}, 
\begin{align}
 \delta_{2} = \delta_{3}  = \frac{\tilde{a}}{\tilde{a}+|\tilde{u}_n|}
 \label{eqn:HLLEM_antidissipationdeltas}
\end{align}
As in the HLLE scheme, we rewrite the HLLEM scheme as, 
\begin{align}
   \mathbf{F}_{HLLEM} = \frac{1}{2}(\mathbf{F}_L + \mathbf{F}_R) + \mathbf{D} + \mathbf{A}
   \label{eqn:HLLEM_centralplusdissipation}
\end{align}
where the dissipation vector $\mathbf{D}$ is same as in Eq.(\ref{eqn:HLLEdissipation}) and represents the embedded HLL-type dissipation while the antidissipation 
vector $\mathbf{A}$ unique to the HLLEM scheme is given as,
\begin{align}
 \mathbf{A} = \frac{S_LS_R}{S_R-S_L}(- \delta_{2} \tilde{\alpha}_{2} \tilde{\mathbf{R}}_{2} - \delta_{3} \tilde{\alpha}_{3} \tilde{\mathbf{R}}_{3})
 \label{eqn:HLLEMantidissipation}
\end{align}
In the above expression, the first term $\delta_{2} \tilde{\alpha}_{2} \tilde{\mathbf{R}}_{2}$ denote antidissipation concerning the contact wave while the second term 
$\delta_{3} \tilde{\alpha}_{3} \tilde{\mathbf{R}}_{3}$ denote that concerning the shear wave. As will be shown in the next section, these terms responsible for 
accuracy on linear
waves will also be responsible for lack of robustness of the HLLEM scheme towards various forms of numerical shock instability.

\section{Instability characteristics of HLLE and HLLEM schemes}
\label{sec:orderofmagnitudeanalysis_hlle_hllem}

\begin{figure}[H]
	    \centering
	    \includegraphics[scale=0.5]{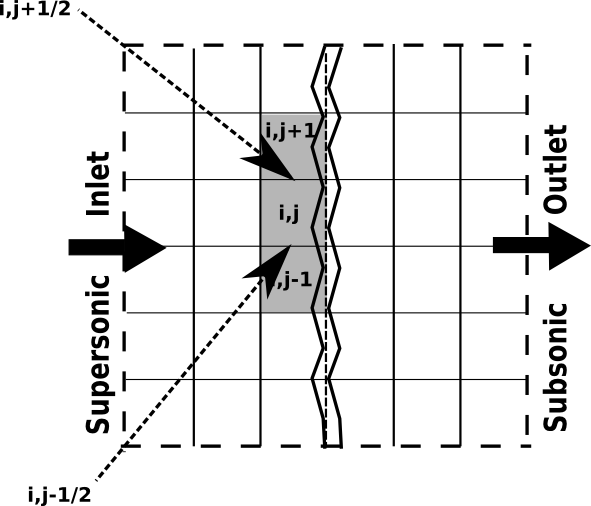}
	    \caption{Schematic showing the stencil chosen for studying the numerical dissipation characterisitcs of the HLLE and the HLLEM schemes.}
	    \label{fig:stencilfororderanalysis}
	    \end{figure}

Studies \cite{shen2014, sangeeth2018_HLLCADC} show that unphysical variation of conserved quantity $\rho u$ along a numerically computed shock front could be a plausible reason for manifestation of shock 
instability. Such unphysical variation could occur independently due to undamped perturbations in density ($\rho$) or flow velocity ($u$) variables \cite{sangeeth2018_HLLCADC}. 
In case of other shock unstable approximate Riemann solvers like the Roe scheme and the HLLC scheme, the proliferation of these perturbations are known to 
occur typically due to reduction in overall numerical dissipation caused by inadvertent activation of their antidiffusive terms on interfaces that are
not aligned with the shock front \cite{shen2014,sangeeth2018_HLLCADC}. Specifically it has been shown in \cite{sangeeth2018_HLLCADC} that activation of antidiffusive 
terms in the mass and interface-normal flux component discretizations 
on these crucial transverse interfaces play a major role in triggering of the instability.
In this context we would like to closely study the behaviour of the numerical dissipation of the shock unstable HLLEM scheme in comparison to that of the shock stable 
HLLE scheme to identify the specific terms of the former that participate in the shock instability mechanism. The study is carried out in a simple setting most commonly known to produce shock instability: 
in the vicinity of a perturbed strong normal shock. For this purpose, consider a y-directional stencil comprising of three candidate cells namely $(i,j), (i,j+1)$ and $(i,j-1)$ located within the upstream coloum of cells 
of an isolated strong normal shock that exists in a steady supersonic flow as shown in Fig.(\ref{fig:stencilfororderanalysis}). 
\noindent The flow is assumed to happen only in positive x direction. Along the transverse direction (ie. along the y-directional interfaces in this case) in the 
vicinity of the shock, the following assumptions can be made without loss of generality:\newline
\begin{align}
 \nonumber
  \rho,u,p \neq 0 \\ 
  v =0 \ \ 
  \label{eqn:assumptionsforupstreamcellsnearshock}
\end{align}

\noindent Our particular concern is the evolution of the conserved quantities $\rho$ and $\rho u$ in cell ($i,j$) due to fluxes that cross the interfaces ($i,j\pm1/2$). 
While the mass fluxes $(\rho v)_{i,j\pm1/2}$ affect the evolution of quantity $\rho$, the interface-normal fluxes  $(\rho u v) _{i,j\pm1/2}$ affect the evolution of
quantity $\rho u$. 
The saw-tooth like perturbations in primitive and conserved variables that are characteristic of a shock unstable solution can be thought to occur due to imbalances in these 
fluxes only \cite{quirk1994}. 
An evolution equation for these quantities can be written as, 

\begin{align}
  \{\rho\}_{i,j}^{n+1} \approx  \{\rho\}_{i,j}^{n} - \frac{\Delta t}{\Delta y} \left [ \{\rho v\}_{i,j+1/2} - \{\rho v\}_{i,j-1/2} \right ]
 \label{eqn:evolutionofdensity}
\end{align}

\begin{align} 
 \{\rho u\}_{i,j}^{n+1} \approx  \{\rho u\}_{i,j}^{n} - \frac{\Delta t}{\Delta y} \left [ \{\rho u v\}_{i,j+1/2} - \{\rho u v\}_{i,j-1/2} \right ]
 \label{eqn:evolutionofmomentum}
\end{align}

\noindent Notice that since there is no primary flow in the y-direction, subsonic fluxes corresponding to the HLLE or HLLEM scheme will be engaged on these interfaces.
Ideally, across these interfaces only information regarding pressure would be transmited through the interface-normal momentum fluxes.
However if numerical perturbations in flow quantities does exist, then additional unphysical fluxes would occur through these interfaces. 
The objective then is to see how the HLLEM scheme differ from the HLLE scheme in its treatment of these unphysical fluxes and the perturbed flow quantities that 
gets convected because of 
them.
In case of the HLLEM scheme, we pay special attention to the role of its antidiffusive terms on the evolution of $\rho$ and $\rho u$ quantities. 

\subsection{Dissipation analysis of the HLLE scheme}
\label{sec:dissipationanalysisofhllescheme}
Suppose consider that we employ the HLLE scheme to evaluate the fluxes in Eqs.(\ref{eqn:evolutionofdensity}) and (\ref{eqn:evolutionofmomentum}).
Then, we are interested in how the numerical dissipation available in the HLLE discretization of these critical flux components affect the evolution of 
$\rho$ and $\rho u$ respectively. Below, we seperately analyze each of them.

\subsubsection{Mass flux}
\label{sec:dissipationanalysisofhllescheme_massflux}
A HLLE discretization for $\{\rho v\}_{i,j+1/2}$ and $\{\rho v\}_{i,j-1/2}$ can be written as,

\begin{align}
 \nonumber
 \{\rho v\}_{j+1/2} = \frac{1}{2} \left[\{\rho v\}_j + \{\rho v\}_{j+1}  \right] + \frac{S_R + S_L}{2(S_R-S_L)} \left[\{\rho v\}_j - \{\rho v\}_{j+1}  \right] 
 -\frac{S_LS_R}{S_R-S_L} \left[ \{\rho \}_{j} - \{\rho \}_{j+1} \right]
\end{align}

\begin{align}
 \nonumber
 \{\rho v\}_{j-1/2} = \frac{1}{2} \left[\{\rho v\}_{j-1} + \{\rho v\}_{j}  \right] + \frac{S_R + S_L}{2(S_R-S_L)} \left[\{\rho v\}_{j-1} - \{\rho v\}_{j}  \right] 
 -\frac{S_LS_R}{S_R-S_L} \left[ \{\rho \}_{j-1} - \{\rho \}_{j} \right]
\end{align}
where we have dropped the $i$ from the interface index for convenience. The flux difference is then, 

\begin{align}
 \nonumber 
  \{\rho v\}_{j+1/2} - \{\rho v\}_{j-1/2} = \frac{1}{2} \left[\{\rho v\}_{j+1} - \{\rho v\}_{j-1}  \right] + \frac{S_R + S_L}{2(S_R-S_L)} \left[\{\rho v\}_{j} - \{\rho v\}_{j+1} 
  -\left( \{\rho v\}_{j-1} - \{\rho u v\}_{j} \right)  \right]\\ \nonumber
  -\frac{S_LS_R}{S_R-S_L} \left[ \{\rho \}_{j}-\{\rho \}_{j+1} - \left( \{\rho u \}_{j-1} - \{\rho \}_{j} \right) \right]
\end{align}
Since our interest lies in contrasting the numerical dissipation behaviours of the HLLE and the HLLEM schemes, it suffices to only consider the dissipation component of the 
flux difference which will be henceforth denoted as ${\Delta D}^{\{\rho v\}_{j+1/2} - \{\rho v\}_{j-1/2}}_{HLLE}$. Using the notation $\Delta(.) = (.)_R - (.)_L$, 
the dissipation component of the above flux difference can be written as, 
\begin{align}
  {\Delta D}^{\{\rho v\}_{j+1/2} - \{\rho v\}_{j-1/2}}_{HLLE} = - \frac{S_R + S_L}{2(S_R-S_L)} \left[ \Delta \{\rho v\}_{j+1/2} - \Delta \{\rho v\}_{j-1/2}  \right]
  +\frac{S_LS_R}{S_R-S_L} \left[ \Delta \{\rho \}_{j+1/2} - \Delta \{\rho \}_{j-1/2} \right]
  \label{eqn:dissipationofmassfluxdifferences_hll}
\end{align}
To introduce the effect of small numerical perturbations that are thought to eventually result in shock unstable solutions, consider the existence of a 
random numerical perturbations of the order 
$\delta$ (eg. due to round of errors) in the flow variables that exists during the course of computation in the stencil considered. 
Then, the following approximations are assumed to hold true on the stencil,

\begin{align}
 \Delta \rho, \Delta u, \Delta v, \Delta p, \Delta (\rho u) \sim \mathcal{O}({\delta}) 
 \label{eqn:assumptionsforupstreamcellswithperturbations}
 \end{align}
The term $\Delta \{\rho v\}_{j\pm1/2}$ can be expanded as $ \tilde{\rho} \Delta v + \tilde{v} \Delta \rho$. Under the above assumptions $\Delta \{\rho v\}_{j\pm1/2}$ is $\mathcal{O}(\delta)$. Assuming $S_L=(\tilde{v}-\tilde{a}) \sim \mathcal{O}({\tilde{a}})$ and 
 $S_R =(\tilde{v}+\tilde{a}) \sim \mathcal{O}({\tilde{a}})$, 
 using Eq.(\ref{eqn:assumptionsforupstreamcellsnearshock}) 
 and Eq.(\ref{eqn:assumptionsforupstreamcellswithperturbations}), Eq.(\ref{eqn:dissipationofmassfluxdifferences_hll}) can be simplified as, 

 \begin{align}
  \Delta D^{\{\rho v\}_{j+1/2} - \{\rho v\}_{j-1/2}}_{HLLE} = - \frac{\cancelto{\mathcal{O}({\delta})}{\tilde{v}}}{2\tilde{a}} \left[ \cancelto{\mathcal{O}({\delta})}{\Delta \{\rho v\}_{j+1/2}} -\cancelto{\mathcal{O}({\delta})}{ \Delta \{\rho v\}_{j-1/2}}  \right]
  \ \ \ \ -\frac{\tilde{a}}{2} \left[ \cancelto{\mathcal{O}({\delta})}{\Delta \{\rho \}_{j+1/2}} - \cancelto{\mathcal{O}({\delta})}{\Delta \{\rho \}_{j-1/2}} \right]
  \label{eqn:dissipationofmassfluxdifferences_hll_simplified1}
\end{align}
Effectively, 
\begin{align} 
  \Delta D^{\{\rho v\}_{j+1/2} - \{\rho v\}_{j-1/2}}_{HLLE} \sim \mathcal{O}(\delta)  
  \label{eqn:dissipationofmassfluxdifferences_hll_simplified2}
\end{align}
Thus in the presence of perturbations of $\mathcal{O}(\delta)$ in flow quantities, the HLLE scheme infuses a net dissipation of the same order into the mass flux 
discretization. 
\subsubsection{Interface-normal momentum flux}
\label{sec:dissipationanalysisofhllescheme_momentumflux}
A HLLE discretization for $\{\rho u v\}_{j+1/2}$ and $\{\rho u v\}_{j-1/2}$ can be written as,

\begin{align}
 \nonumber
 \{\rho u v\}_{j+1/2} = \frac{1}{2} \left[\{\rho u v\}_j + \{\rho u v\}_{j+1}  \right] + \frac{S_R + S_L}{2(S_R-S_L)} \left[\{\rho u v\}_j - \{\rho u v\}_{j+1}  \right] 
 -\frac{S_LS_R}{S_R-S_L} \left[ \{\rho u \}_{j} - \{\rho u \}_{j+1} \right]
\end{align}

\begin{align}
 \nonumber
 \{\rho u v\}_{j-1/2} = \frac{1}{2} \left[\{\rho u v\}_{j-1} + \{\rho u v\}_{j}  \right] + \frac{S_R + S_L}{2(S_R-S_L)} \left[\{\rho u v\}_{j-1} - \{\rho u v\}_{j}  \right] 
 -\frac{S_LS_R}{S_R-S_L} \left[ \{\rho u \}_{j-1} - \{\rho u \}_{j} \right]
\end{align}
%
%

\noindent The dissipation component of the flux difference in this case denoted as $\Delta D^{\{\rho u v\}_{j+1/2} - \{\rho u v\}_{j-1/2}}_{HLLE}$ will be, 
\begin{align}
  \Delta D^{\{\rho u v\}_{j+1/2} - \{\rho u v\}_{j-1/2}}_{HLLE} = - \frac{S_R + S_L}{2(S_R-S_L)} \left[ \Delta \{\rho u v\}_{j+1/2} - \Delta \{\rho u v\}_{j-1/2}  \right]
  +\frac{S_LS_R}{S_R-S_L} \left[ \Delta \{\rho u \}_{j+1/2} - \Delta \{\rho u \}_{j-1/2} \right]
  \label{eqn:dissipationofmomentumfluxdifferences_hll}
\end{align}
 The term $\Delta \{\rho u v\}_{j\pm1/2}$ can be expanded in terms of the mass flux as $ \tilde{\rho} \tilde{u} \Delta v + \tilde{v} \Delta(\rho u) $ and under the 
 assumptions made in Eq.(\ref{eqn:assumptionsforupstreamcellswithperturbations}) 
 is $\mathcal{O}(\delta)$. Using Eq.(\ref{eqn:assumptionsforupstreamcellsnearshock}) and Eq.(\ref{eqn:assumptionsforupstreamcellswithperturbations}), 
 Eq.(\ref{eqn:dissipationofmomentumfluxdifferences_hll}) can be simplified as, 
 \allowbreak
\begin{align}
  \Delta D^{\{\rho u v\}_{j+1/2} - \{\rho u v\}_{j-1/2}}_{HLLE} = - \frac{ \cancelto{\mathcal{O}({\delta})}{\tilde{v}}}{2\tilde{a}} \left[ \cancelto{\mathcal{O}({\delta})}{ \Bigl({\tilde{\rho} \tilde{u} \Delta v + \tilde{v} \Delta(\rho u)}\Bigr)_{j+1/2}} -\cancelto{\mathcal{O}({\delta})}{ \Bigl({\tilde{\rho} \tilde{u} \Delta v + \tilde{v} \Delta(\rho u)}\Bigr)_{j-1/2}}  \right]
  \ \ \ -\frac{\tilde{a}}{2} \left[ \cancelto{\mathcal{O}({\delta})}{\Delta \{\rho u \}_{j+1/2}} - \cancelto{\mathcal{O}({\delta})}{\Delta \{\rho u \}_{j-1/2}} \right]
  \label{eqn:dissipationofmoemtumfluxdifferences_hll_simplified1}
\end{align}
Which indicates, 
\begin{align}
  \Delta D^{\{\rho u v\}_{j+1/2} - \{\rho u v\}_{j-1/2}}_{HLLE} \sim \mathcal{O}(\delta)  
  \label{eqn:dissipationofmomentumfluxdifferences_hll_simplified2}
\end{align}
Thus in the presence of perturbations of $\mathcal{O}(\delta)$ in flow quantities, the HLLE scheme infuses a net dissipation of the same order into the discretization of 
interface-normal momentum flux in the y-direction. 

\subsubsection{Role of the activated diffusive terms in the HLLE scheme}
\label{sec:roleofHLLEdissipation}
\begin{figure}[H]
	    \setcounter{subfigure}{0}
	    \subfloat[ $\mathbf{HLLE}$ $\boldsymbol{\left(\hat{\rho} = 0.01, \hat{u} = 0, \hat{p} = 0\right)} $ ]{\label{fig:HLLE_density_perturbation}\includegraphics[scale=0.30]{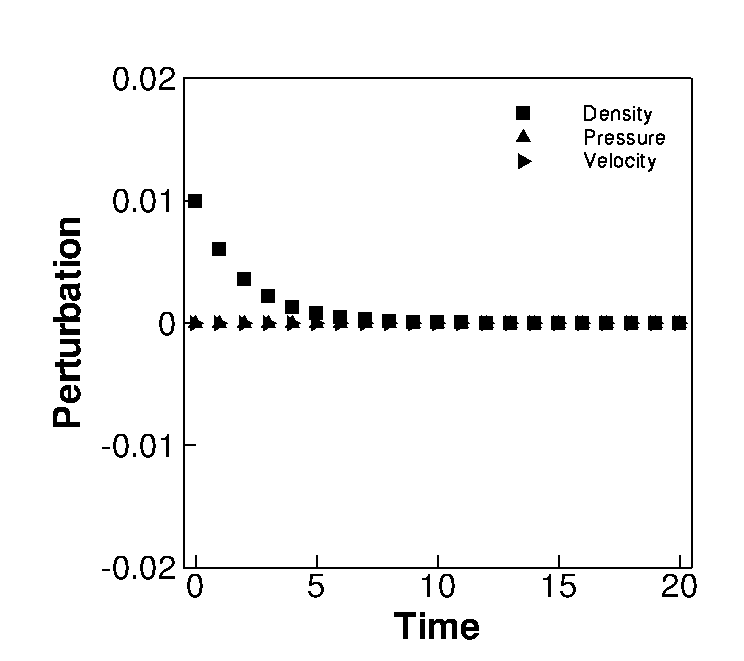}}
            \qquad
            \qquad
            \subfloat[$\mathbf{HLLE}$ $\boldsymbol{\left( \hat{\rho} = 0, \hat{u} = 0.01, \hat{p} = 0 \right) }$]{\label{fig:HLLE_velocity_perturbation}\includegraphics[scale=0.30]{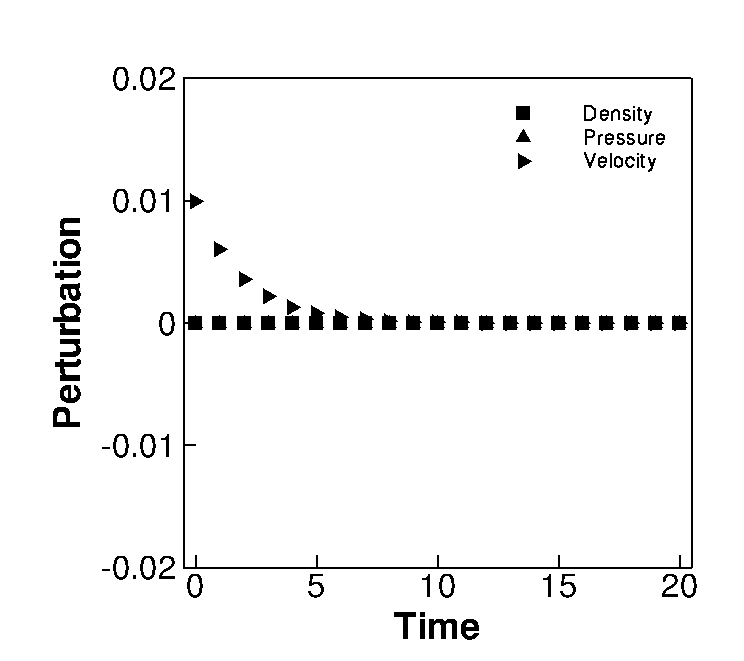}}\\
            \subfloat[$\mathbf{HLLE}$ $\boldsymbol{ \left(\hat{\rho} = 0, \hat{u} = 0, \hat{p} = 0.01\right) }$]{\label{fig:HLLE_pressure_perturbation}\includegraphics[scale=0.30]{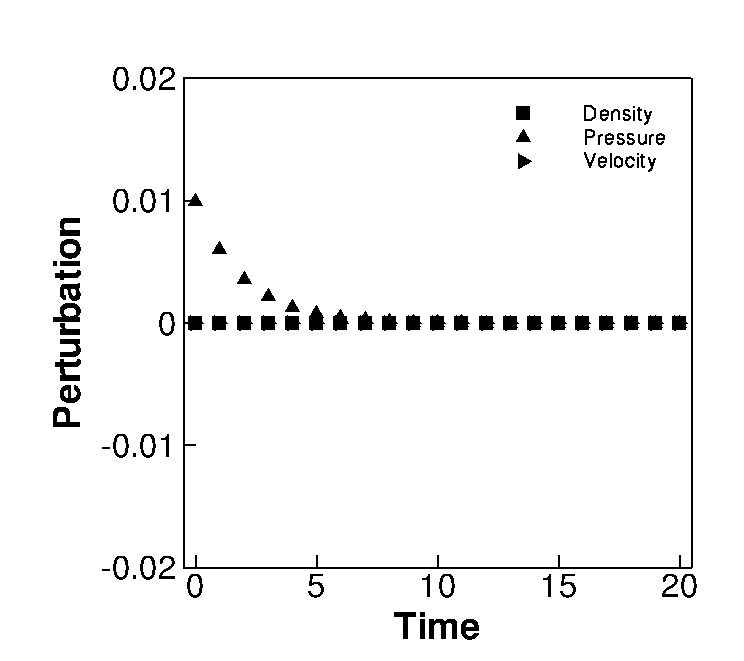}} 
	    \caption{Comparison of evolution of density, x-velocity and pressure perturbations in the HLLE scheme.}
	    \label{fig:perturbationstudies_HLLE}
	    \end{figure}

It was seen in Sections.(\ref{sec:dissipationanalysisofhllescheme_massflux}) and (\ref{sec:dissipationanalysisofhllescheme_momentumflux}) that a numerical perturbation in the
flow quantities, in the vicinity of a normal shock, tends to activate the dissipation component of the HLLE scheme in both mass and interface-normal momentum flux components.
To confirm whether this amount of dissipation suffices to damp out the perturbations in flow variables like $\rho$, $u$ and $p$ which affect the quantities $\rho$ and 
$\rho u$, we resort to a technique of linear perturbation analysis. Mainly, we seek to develop the linearized evolution equations for 
perturbation in these variables from the approximate evolution equations given in Eqs.(\ref{eqn:evolutionofdensity}) and (\ref{eqn:evolutionofmomentum}) as per the technique 
provided in \cite{pandolfi2001}. 
For developing the evolution equation for $p$ we additionally utilize an approximate evolution equation of the total energy in transverse direction given by, 
\begin{align}
 \{\rho E\}_{i,j}^{n+1} \approx  \{\rho E\}_{i,j}^{n} - \frac{\Delta t}{\Delta y} \left [ \{(\rho E +p) v\}_{i,j+1/2} - \{(\rho E + p) v\}_{i,j-1/2} \right ]
 \label{eqn:evolutionofenergy}
\end{align}
In the present work, the steady mean flow is chosen to have normalised 
state values of $\rho_0=1$, $u_0\neq0$, $v_0=0$ and $p_0=1$. The perturbations in each of these variables denoted as $\hat{\rho}, \hat{u}$ and $\hat{p}$ 
are introduced as a saw tooth profile superimposed on the steady mean flow. Thus for a typical cell 'j', the perturbed initial conditions are,
\begin{align}
\begin{cases}
    \rho_j=\rho_0+\hat{\rho}, \;\;p_j=p_0+\hat{p},\;\;u_j=u_0+\hat{u}\;\;& \text{if } j\;\; is\;\; even\\
    \rho_j=\rho_0-\hat{\rho}, \;\;p_j=p_0-\hat{p},\;\;u_j=u_0-\hat{u}\;\;& \text{if } j\;\; is\;\; odd
\end{cases}
\label{eqn:sawtoothinitialconditions}
\end{align}
Evolution equations for perturbations in primitive variables $\rho$, $u$ and $p$ in case of the HLLE scheme are given in Table \ref{table:evolutionequations} where 
$\lambda = \sqrt{\gamma}\frac{\Delta t}{\Delta y}$ denotes a linearized CFL value. The amplification factors for density, x-velocity and pressure in case of HLLE scheme are 
$(1-2\lambda,1-2\lambda,1-2\lambda)$ respectively. Thus for $0<\lambda<1$, any initial perturbation in these flow variables would be effectively damped by the HLLE dissipation.   
The evolution of these perturbations are plotted in Fig.(\ref{fig:perturbationstudies_HLLE}) and is more illustrative of the above discussion.  
Each plot in Fig.(\ref{fig:perturbationstudies_HLLE})
indicate the evolution of all three perturbations: $\hat{\rho}, \hat{u}$ and $\hat{p}$ for a 
given initial perturbation in only one of those quantities (while setting the other perturbations to be 0). For all the experiments $\lambda$ is taken to be 0.2.


\sloppy 
 
 \begin{table}[H]
    \caption{Evolution equations for perturbations in primitive variables for various schemes} 
    \centering 
    \begin{tabular}{|c|c|c|c|} 
    \hline 
    Scheme & Density & Pressure & x-velocity\\ [0.3ex] 
    \hline 
    HLLE & $\hat{\rho}^{n+1}=(1- 2 \lambda) \hat{\rho}^n$ & $\hat{P}^{n+1}= \hat{P}^n(1-2 \lambda) $&  $\hat{u}^{n+1}= (1- 2 \lambda)\hat{u}^n $ \\ 
    HLLEM & $\hat{\rho}^{n+1}= \hat{\rho}^n - \frac{2 \lambda}{\gamma}\hat{P}^n$ & $\hat{P}^{n+1}= \hat{P}^n(1-2 \lambda) $ & $\hat{u}^{n+1}= \hat{u}^n $ \\
    HLLEM-SWM-P/E &  $\hat{\rho}^{n+1}= (1- \frac{2\lambda\alpha\epsilon}{\sqrt{\gamma}})\hat{\rho}^n - \frac{2 \lambda}{\gamma}\hat{P}^n$ & $\hat{P}^{n+1}= \hat{P}^n(1-2\lambda(1+\frac{\alpha\epsilon}{\sqrt{\gamma}})) $ & $\hat{u}^{n+1}= \hat{u}^n(1-\frac{2\lambda\alpha\epsilon}{\sqrt{\gamma}}) $\\
    HLLEM-ADC &  $\hat{\rho}^{n+1}= (1-2\lambda(1-\omega))\hat{\rho}^n - \frac{2\omega \lambda}{\gamma}\hat{P}^n$ & $\hat{P}^{n+1}= \hat{P}^n(1-2 \lambda) $ & $\hat{u}^{n+1}= \hat{u}^n(1-2\lambda(1-\omega)) $\\
    HLLEMS &  $\hat{\rho}^{n+1}= \hat{\rho}^n - \frac{2 \lambda}{\gamma}\hat{P}^n$ & $\hat{P}^{n+1}= \hat{P}^n(1-2 \lambda) $ & $\hat{u}^{n+1}= \hat{u}^n(1-2\lambda(1-\omega)) $\\
    \hline 
    \end{tabular}
    \label{table:evolutionequations} 
    \end{table}

\noindent From the behaviour of these evolution equations, it is clear that any perturbation in all three flow variables are attenuated in time by 
the HLLE scheme. This indicates that the activated dissipative terms of the HLLE scheme as found in Eqs.(\ref{eqn:dissipationofmassfluxdifferences_hll_simplified2}) and 
(\ref{eqn:dissipationofmomentumfluxdifferences_hll_simplified2}) are capable of damping any unwanted perturbations that may exist in the course of a numerical computation. This in effect ensures that 
variation of the conserved quantity $\rho u$ along the shock front remains spatially and temporally unvarying and the shock structure is preserved through an accurate Rankine-Hugoniot jump across it. 

\subsection{Dissipation analysis of the HLLEM scheme}
\label{sec:dissipationanalysisofhllemscheme}

Consider that instead of the HLLE scheme, we use the HLLEM scheme to evaluate the mass and interface-normal momentum fluxes in in Eqs.(\ref{eqn:evolutionofdensity}) and (\ref{eqn:evolutionofmomentum})
at these crucial transverse interfaces.
Then, we are interested in how the numerical dissipation provided by the HLLEM scheme differ from that already seen in the HLLE scheme.
Below, we seperately analyze each of them.
Since HLLEM scheme can be thought of as a combination of the HLLE scheme and antidiffusive terms that are responsible for accuracy on 
linear waves, we specially study the role of these antidiffusive terms in dealing with these perturbations.

\subsubsection{Mass flux}
\label{sec:dissipationanalysisofhllemscheme_massflux}

A HLLEM discretization for $\{\rho v\}_{i,j+1/2}$ and $\{\rho v\}_{i,j-1/2}$ can be written as,
\sloppy 

\begin{align}
 \{\rho v\}_{j+1/2} = \frac{1}{2} \left[\{\rho v\}_j + \{\rho v\}_{j+1}  \right] + \frac{S_R + S_L}{2(S_R-S_L)} \left[\{\rho v\}_j - \{\rho v\}_{j+1}  \right] 
 -\frac{S_LS_R}{S_R-S_L} \left[ \{\rho \}_{j} - \{\rho \}_{j+1} \right] + \frac{S_LS_R}{S_R-S_L} \left[ \{-\delta_2 \tilde{\alpha_2}\}_{j+1/2} \right]
\end{align}

\begin{align}
 \{\rho v\}_{j-1/2} = \frac{1}{2} \left[\{\rho v\}_{j-1} + \{\rho v\}_{j}  \right] + \frac{S_R + S_L}{2(S_R-S_L)} \left[\{\rho v\}_{j-1} - \{\rho v\}_{j}  \right] 
 -\frac{S_LS_R}{S_R-S_L} \left[ \{\rho \}_{j-1} - \{\rho \}_{j} \right] + \frac{S_LS_R}{S_R-S_L} \left[ \{-\delta_2 \tilde{\alpha_2}\}_{j-1/2} \right]
\end{align}

\noindent The flux difference is then, 

\begin{align}
  \{\rho v\}_{j+1/2} - \{\rho v\}_{j-1/2} = \frac{1}{2} \left[\{\rho v\}_{j+1} - \{\rho v\}_{j-1}  \right] + \frac{S_R + S_L}{2(S_R-S_L)} \left[\{\rho v\}_{j} - \{\rho v\}_{j+1} 
  -\left( \{\rho v\}_{j-1} - \{\rho u v\}_{j} \right)  \right]\\ \nonumber
  -\frac{S_LS_R}{S_R-S_L} \left[ \{\rho \}_{j}-\{\rho \}_{j+1} - \left( \{\rho u \}_{j-1} - \{\rho \}_{j} \right) \right] + \frac{S_LS_R}{S_R-S_L} \left[\{-\delta_2 \tilde{\alpha_2}\}_{j+1/2} 
  +  \{\delta_2 \tilde{\alpha_2}\}_{j-1/2} \right]
\end{align}

\noindent Since we are interested only in the numerical dissipation provided by the HLLEM scheme, we seek to isolate these terms from the above equation. 
These terms can be extracted as, 
\begin{align}
  {\Delta D}^{\{\rho v\}_{j+1/2} - \{\rho v\}_{j-1/2}}_{HLLEM} = {\Delta D}^{\{\rho v\}_{j+1/2} - \{\rho v\}_{j-1/2}}_{HLLE} + \frac{S_LS_R}{S_R-S_L} \left[\{-\delta_2 \tilde{\alpha_2}\}_{j+1/2} 
  +  \{\delta_2 \tilde{\alpha_2}\}_{j-1/2} \right]
  \label{eqn:dissipationofmassfluxdifferences_hllem}
  \end{align}
Notice that we have expressed the numerical dissipation of the HLLEM scheme as a combination of HLL-type dissipation (denoted by ${\Delta D}^{\{\rho v\}_{j+1/2} - \{\rho v\}_{j-1/2}}_{HLLE}$) 
and dissipation arising from its antidiffusive terms. 
An order of magnitude estimate for ${\Delta D}^{\{\rho v\}_{j+1/2} - \{\rho v\}_{j-1/2}}_{HLLE}$ has been 
already presented in Eq.(\ref{eqn:dissipationofmassfluxdifferences_hll_simplified2}).
We refer to the second term that arises from antidiffusive component of the HLLEM scheme as ${\Delta A}^{\{\rho v\}_{j+1/2} - \{\rho v\}_{j-1/2}}_{HLLEM}$. Now we seek an 
estimate for this term. Firstly notice that using the conditions presented in Eq.(\ref{eqn:assumptionsforupstreamcellsnearshock})and 
Eq.(\ref{eqn:assumptionsforupstreamcellswithperturbations}), the wave strength $\tilde{\alpha_2}$ and antidissipation coefficient $\delta_2$ at each interface
can be simplified as, 

\begin{align}
\nonumber
\tilde{\alpha_2} =  \cancelto {\mathcal{O}({\delta})}{\Delta \rho} \ \ - \ \ \frac{ \cancelto {\mathcal{O}({\delta})} {\Delta p}}{\tilde{a}^2} \sim \mathcal{O}({\delta}) \\
\delta_2 =  \frac{\tilde{a}}{\tilde{a}+|\tilde{v}|} \sim \mathcal{O}({1})
\end{align}
Using these the term ${\Delta A}^{\{\rho v\}_{j+1/2} - \{\rho v\}_{j-1/2}}_{HLLEM}$ can be estimated to be, 
\begin{align}
 {\Delta A}^{\{\rho v\}_{j+1/2} - \{\rho v\}_{j-1/2}}_{HLLEM} \sim \mathcal{O}(\delta) 
 \label{eqn:antidissipationofmassfluxdifferences_hllem_simplified}
\end{align}

\noindent Eq.(\ref{eqn:antidissipationofmassfluxdifferences_hllem_simplified}) reveals something interesting. In the presence of numerical perturbations of $\mathcal{O}(\delta)$, it is observed that the 
antidiffusive terms in the HLLEM massflux discretization is activated and is also of $\mathcal{O}(\delta)$ similar to its inherent HLL-type diffusive terms. 

\subsubsection{Interface-normal momentum flux}
\label{sec:dissipationanalysisofhllemscheme_momentumflux}
A HLLEM discretization for $\{\rho u v\}_{j+1/2}$ and $\{\rho u v\}_{j-1/2}$  can be written as, 

\begin{align}
 \nonumber
 \{\rho u v\}_{j+1/2} = \frac{1}{2} \left[\{\rho u v\}_j + \{\rho u v\}_{j+1}  \right] + \frac{S_R + S_L}{2(S_R-S_L)} \left[\{\rho u v\}_j - \{\rho u v\}_{j+1}  \right] 
 -\frac{S_LS_R}{S_R-S_L} \left[ \{\rho u \}_{j} - \{\rho u \}_{j+1} \right]  \\ 
 + \frac{S_LS_R}{S_R-S_L} \left[ \{-\delta_2 \tilde{\alpha_2} \tilde{u} -\delta_3 \tilde{\alpha_3}\}_{j+1/2} \right]
\end{align}
and, 

\begin{align}
 \nonumber
 \{\rho u v\}_{j-1/2} = \frac{1}{2} \left[\{\rho u v\}_{j-1} + \{\rho u v\}_{j}  \right] + \frac{S_R + S_L}{2(S_R-S_L)} \left[\{\rho u v\}_{j-1} - \{\rho u v\}_{j}  \right] 
 -\frac{S_LS_R}{S_R-S_L} \left[ \{\rho u \}_{j-1} - \{\rho u \}_{j} \right] \\ 
 + \frac{S_LS_R}{S_R-S_L} \left[ \{-\delta_2 \tilde{\alpha_2} \tilde{u} -\delta_3 \tilde{\alpha_3}\}_{j-1/2}  \right]
\end{align}

\noindent Flux difference in y-direction is then, 

\begin{align}
 \nonumber 
  \{\rho u v\}_{j+1/2} - \{\rho u v\}_{j-1/2} = \frac{1}{2} \left[\{\rho u v\}_{j+1} + \{\rho u v\}_{j-1}  \right] + \frac{S_R + S_L}{2(S_R-S_L)} \left[\{\rho u v\}_{j} - \{\rho u v\}_{j+1} 
  -\{\rho u v\}_{j-1} + \{\rho u v\}_{j}  \right]\\ \nonumber
  -\frac{S_LS_R}{S_R-S_L} \left[ \{\rho u \}_{j}-\{\rho u \}_{j+1} + \{\rho u \}_{j} - \{\rho u \}_{j-1} \right] + \frac{S_LS_R}{S_R-S_L} \left[ \{ -\delta_2 \tilde{\alpha_2} \tilde{u} -\delta_3 \tilde{\alpha_3}\}_{j+1/2} 
  +  \{\delta_2 \tilde{\alpha_2} \tilde{u} +\delta_3 \tilde{\alpha_3}\}_{j-1/2}\right]
\end{align}

\noindent The terms corresponding to numerical dissipation can be extracted as, 
\begin{align}
  {\Delta D}^{\{\rho u v\}_{j+1/2} - \{\rho u v\}_{j-1/2}}_{HLLEM} = {\Delta D}^{\{\rho u v\}_{j+1/2} - \{\rho u v\}_{j-1/2}}_{HLLE} + \frac{S_LS_R}{S_R-S_L} \left[ \{ -\delta_2 \tilde{\alpha_2} \tilde{u} -\delta_3 \tilde{\alpha_3}\}_{j+1/2} 
  +  \{\delta_2 \tilde{\alpha_2} \tilde{u} +\delta_3 \tilde{\alpha_3}\}_{j-1/2}\right]
  \label{eqn:dissipationoffluxdifferences_hllem}
\end{align}
We have again expressed the numerical dissipation of the HLLEM scheme as a combination of a HLL-type dissipation (denoted by ${\Delta D}^{\{\rho u v\}_{j+1/2} - \{\rho u v\}_{j-1/2}}_{HLLE}$) 
and dissipation arising from the antidiffusive terms. 
An order of magnitude estimate for ${\Delta D}^{\{\rho u v\}_{j+1/2} - \{\rho u v\}_{j-1/2}}_{HLLE}$ has been 
already presented in Eq.(\ref{eqn:dissipationofmomentumfluxdifferences_hll_simplified2}).
We refer to the second term that arises from the antidiffusive terms of the HLLEM scheme as ${\Delta A}^{\{\rho u v\}_{j+1/2} - \{\rho u v\}_{j-1/2}}_{HLLEM}$.
By using the conditions presented in Eq.(\ref{eqn:assumptionsforupstreamcellsnearshock})and Eq.(\ref{eqn:assumptionsforupstreamcellswithperturbations}), this 
term can be further simplified. Note that under the prescribed conditions, the wave strength $\tilde{\alpha_3} \sim \mathcal{O}(\delta)$. Hence it can be shown that,

\begin{align}
 {\Delta A}^{\{\rho u v\}_{j+1/2} - \{\rho u v\}_{j-1/2}}_{HLLEM} \sim \mathcal{O}(\delta) 
 \label{eqn:antidissipationofmomentumfluxdifferences_hllem_simplified}
\end{align}
Once again, it is observed that numerical perturbations of $\mathcal{O}(\delta)$ have activated the antidiffusive terms in the HLLEM interface-normal momentum 
flux discretization in y-direction. Noticeably, this term is also of $\mathcal{O}(\delta)$ similar to its inherent HLL-type diffusive terms.

\subsubsection{Role of the activated antidiffusive terms in the HLLEM scheme}
\label{sec:roleofHLLEMantidissipation}
\begin{figure}[H]
	    \setcounter{subfigure}{0}
            \subfloat[$\mathbf{HLLEM}$ $ \boldsymbol{\left(\hat{\rho} = 0.01, \hat{u} = 0, \hat{p} = 0\right) } $]{\label{fig:HLLEM_density_perturbation}\includegraphics[scale=0.30]{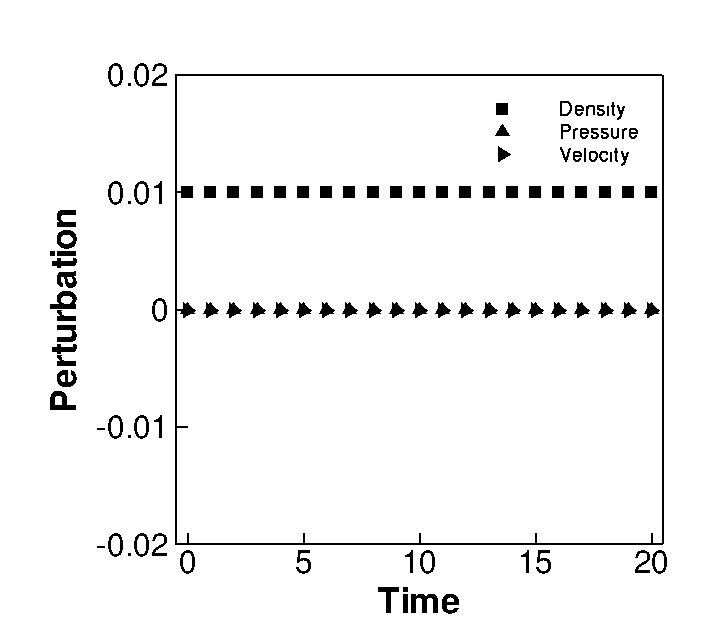}}
            \qquad
            \qquad
            \subfloat[$\mathbf{HLLEM}$  $\boldsymbol{\left(\hat{\rho} = 0, \hat{u} = 0.01, \hat{p} = 0\right) }$]{\label{fig:HLLEM_velocity_perturbation}\includegraphics[scale=0.30]{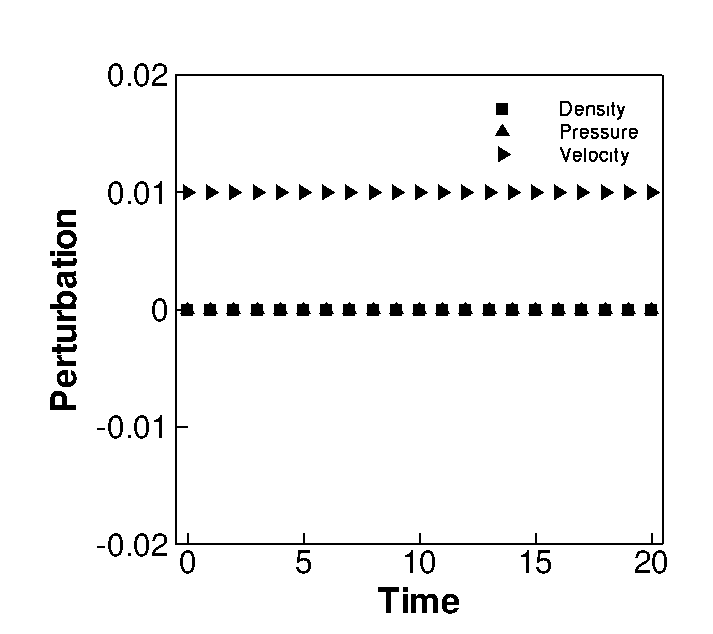}}\\
            \subfloat[$\mathbf{HLLEM}$ $\boldsymbol{ \left(\hat{\rho} = 0, \hat{u} = 0, \hat{p} = 0.01\right) } $]{\label{fig:HLLEM_pressure_perturbation}\includegraphics[scale=0.30]{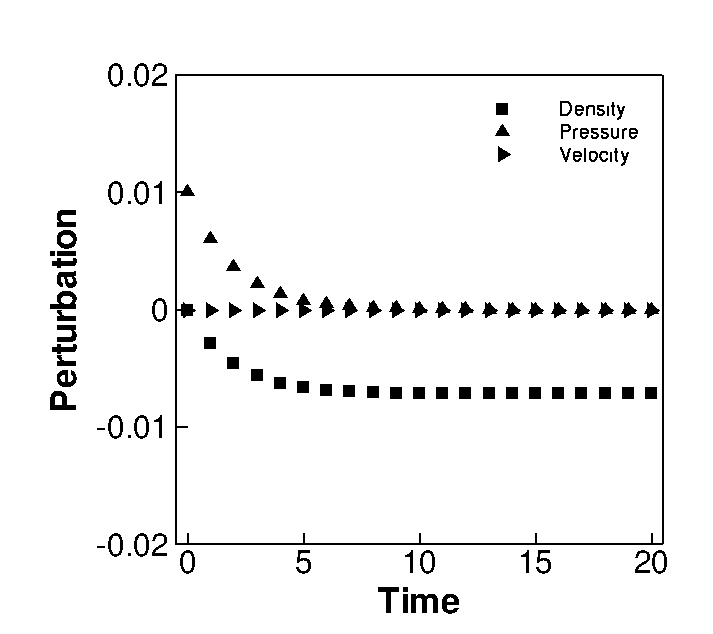}}
	    \caption{Comparison of evolution of density, x-velocity and pressure perturbations in HLLEM scheme.}
	    \label{fig:perturbationstudies_HLLEM}
	    \end{figure}

To see the effect of these antidiffusive terms more clearly, we develop the corresponding linearized evolution equations for perturbations in $\rho$, $u$ and $p$
for the HLLEM scheme similar to the ones described for the HLLE scheme in section(\ref{sec:roleofHLLEdissipation}). These are given in Table \ref{table:evolutionequations}.
It is seen from these equations that in case of the HLLEM scheme, the amplification factors for perturbations in $\rho$, $u$ and $p$ are $(1,1,1-2\lambda)$ respectively. 
This indicates that
the HLLEM scheme is incapable of damping out perturbations in $\rho$ and $u$ variables. 
Additionally it is seen that although there exist a mechanism to damp out perturbations in $p$ for $0<\lambda<1$, any finite $\hat{p}$ is  
repeatedly fed into $\hat{\rho}$ which results in the existence of a residual $\hat{\rho}$ that is growing with time. The behaviour of these evolution equations
are given in Fig.(\ref{fig:perturbationstudies_HLLEM}).
To specifically understand the role of antidiffusive terms in triggering shock instability, we rewrite the evolution equations for perturbations in $\rho$ and $u$ variables as,

  \begin{align}
      \nonumber
      \hat{\rho}^{n+1}&= \underbrace{\hat{\rho}^n (1-2\lambda)}_{HLL \ component} +  \  \underbrace{ (2\lambda(\hat{\rho}^n - \frac{\hat{p}^n}{\gamma}))}_{Antidiffusive \ component} \\\nonumber
      \hat{u}^{n+1} &= \underbrace{\hat{u}^n (1-2\lambda)}_{HLL \ component} + \  \underbrace{ (2\lambda \hat{u}^n)}_{Antidiffusive \ component}\\
      \label{eqn:quirk_analysis_rearranged_hllem}
      \end{align}
      
\noindent In these evolution equations,  we have clearly distinguished the contribution from the inherent HLL-type diffusive terms and the antidiffusive terms. 
Suppose during a simulation, random numerical perturbations of $\mathcal{O} (\delta)$ exist in flow variables along a strong normal shock wave.
From Table (\ref{table:evolutionequations}) we know that the HLLEM scheme has an inherent damping mechanism specifically for pressure perturbations. 
However, as seen from Eqs.(\ref{eqn:antidissipationofmassfluxdifferences_hllem_simplified}) and (\ref{eqn:antidissipationofmomentumfluxdifferences_hllem_simplified}) 
these perturbations inadvertently
activate the antidiffusive terms in the mass flux component and interface-normal momentum flux component on the interfaces transverse to the shock front. These activated terms
are also of $\mathcal{O} (\delta)$.
The erroneously activated antidiffusive terms introduce additional density 
and pressure perturbations into the density evolution equation (obtained from mass flux equation) and extra x-velocity perturbation into the x-velocity perturbation evolution
equation (obtained from x-momentum equation). These additional terms counteract the dissipative action
of the inherently present HLL terms and causes growth in the $\hat{\rho}$ and $\hat{u}$ perturbations. In fact by their very construction, the antidiffusive terms of the HLLEM scheme are designed to reduce the dissipation 
associated with its inherent HLL two wave approximation and provide accuracy on linear wavefields. 
It is known that undamped perturbations in $\hat{\rho}$ or $\hat{u}$ can independently
cause undesirable variations in the conserved quantity $\rho u$ along the shock front \cite{sangeeth2018_HLLCADC}. Such errors could propagate into other equations due to the 
nonlinear coupling nature of the Euler equations. These unphysical variations could then be advected into the numerical shock forcing the shock structure to 
adjust through a typical 'bulge' or in worst cases a complete breakdown of it as is observed in a shock unstable solution. Thus we see that the mechanism through which the
instability is triggered in the HLLEM scheme is analogous to that of the HLLC scheme discussed in \cite{sangeeth2018_HLLCADC}. Hence we concieve that cures that were developed for the 
HLLC scheme in \cite{sangeeth2018_HLLCADC,sangeeth2018_HLLCSLSR} could be extended for the HLLEM scheme.
In the following section we explore this possibility and build various versions of shock stable HLLEM scheme.


\section{Shock stable formulations of HLLEM scheme}
\label{sec:formulations}
The analyses in the last section reveals that shock instability in the HLLEM scheme could be triggered due to inadvertent activation of its antidiffusive terms in mass and 
interface-normal momentum flux component discretizations, 
in the transverse direction of a numerically perturbed shock front, which in turn could weaken its inherently available HLL-type diffusive terms on these components. 
Based on this understanding, we develop two strategies to save the HLLEM scheme from this failure.

\subsection{Shock stable HLLEM scheme based on enhancing inherent HLL-type dissipation}
\label{sec:HLLE_SWM_formulation}
Since the antidiffusive terms present in the mass and interface-normal HLLEM flux discretizations in the transverse direction of a shock front 
are chiefly responsible for triggering the instability, a straigtforward method to achieve a shock stable HLLEM scheme could be to weaken these antidiffusive terms on these
crucial interfaces in the vicinity of a shock. We explore this idea further in Section (\ref{sec:HLLEM-ADC_formulation}). However, an interesting alternative to this philosophy 
would be to instead focus on increasing the magnitude of its inherent HLL-type diffusion vector $\mathbf{D}$ on these critical flux components in the vicinity of a shock.
It has been shown in \cite{sangeeth2018_HLLCSLSR} that such an extra dissipation can be introduced into the HLL-based schemes by appropriately 
modifying the nonlinear wavespeed estimates $S_L$ and $S_R$ appearing in the diffusion vector $\mathbf{D}$. This has been shown to counter the adverse effect of 
antidiffusive terms on stability of shock profiles. Hence following correction to the nonlinear wave speed estimates appearing only in the diffusion
$\mathbf{D}$ of
these critical flux components is made as,

\begin{align}
\nonumber
 \overline{S_L} = S_L - \alpha\epsilon^{L}\\
 \overline{S_R} = S_R + \alpha\epsilon^{R}
 \label{eqn:hllem_modifiedwavespeedestimates}
\end{align}
$S_L$ and $S_R$ in the above equations are obtained from Eq.(\ref{eqn:HLLEwavespeedestimate}). The term $\epsilon^{L,R}$ is an appropriate choice of quantity of 
dissipation to be introduced which is
of the order $\mathcal{O} (S_L,S_R)$ and $\alpha$ is a coefficient through which the 
quantity of $\epsilon^{L,R}$ introduced into $S_L, S_R$ can be controlled. Note that the amount of dissipation to $S_L,S_R$ can be varied independently by defining unique
$\epsilon^{L,R}$ for each of them.  Both $\epsilon^{L,R}$ and $\alpha$ are by definition nonnegative quantities. We set $\alpha$ as 3.5 following
conclusions from \cite{sangeeth2018_HLLCSLSR}. Two methods for estimation of $\epsilon^{L,R}$ that are inspired from the physical nature of the flow are discussed in 
Sec.(\ref{sec:characteristicbasedestimation}) and Sec.(\ref{sec:pressurebasedestimation}).  
These modified wave speed estimates can be used to enhance the magnitude of the HLLE-type dissipation by incorporating them into the diffusion vector itself. This is achieved by defining
a modified diffusion vector $\mathbf{\overline{D}}$ for only the mass and interface-normal momentum flux components as, 
\begin{align}
\mathbf{\overline{D}} = \frac{|\overline{S_R}|-|\overline{S_L}|}{2(S_R-S_L)}(\mathbf{F}_L-\mathbf{F}_R) +  \frac{|\overline{S_L}|S_R-|\overline{S_R}|S_L}{2(S_R-S_L)}(\mathbf{U}_L-\mathbf{U}_R)
\label{eqn:HLLEdissipation-absoluteform}
\end{align}
Although experiments in \cite{sangeeth2018_HLLCADC} indicate that treatment of these flux components on the critical transverse interfaces are sufficient to ensure 
shock stability in case of the HLLC scheme, we observed numerically that this is only a necessary but not a sufficient condition for shock stability in case of the 
HLLEM scheme. 
Infact, it was observed that when employing the cure involving the modified diffusion vector $\mathbf{\overline{D}}$, treatment of mass and interface-normal flux 
discretizations on interfaces in both directions in the vicinity of a shock front: 
the ones parallel to it and the ones across it ensured shock stability in case of the HLLEM scheme. This may be because the unphysical fluxes that develop along the 
shock front, due to the unfavourable action of
the antidiffusive terms, gets convected into the 
numerical shock structure at a faster rate than they could be damped through the increased HLL-type dissipation achieved by $\mathbf{\overline{D}}$.
Hence, we recommend employing $\mathbf{\overline{D}}$ on mass and interface-normal flux discretizations on all interfaces in the vicinity 
of the shock. Hence this type of modified HLLEM scheme, henceforth referred to as HLLEM-SWM (\textbf{S}elective \textbf{W}ave \textbf{M}odified)
can be written as, 

\begin{align}
   \mathbf{F}_{HLLEM-SWM} =
\begin{cases}
   \frac{1}{2}(\mathbf{F}_L + \mathbf{F}_R) + \mathbf{\overline{D}} + \mathbf{A}, & \text{if } mass,\;interface-normal\;momentum\\
   \frac{1}{2}(\mathbf{F}_L + \mathbf{F}_R) + \mathbf{D} + \mathbf{A}, & \text{if } energy,\;interface-tangent\;momentum
\end{cases}
   \label{eqn:HLLEMSWMflux}
\end{align}

\noindent Note that the wavespeeds $S_L, S_R$ present in the coefficient of the antidissipation vector $\mathbf{A}$
are left completely unaffected by this modification. This is done purposefully to avoid affecting the linear wave resolution ability of the HLLEM solver. 
In the coming sections,
we discuss two different strategies for estimation of $\epsilon^{L,R}$ parameter.

\subsubsection{Characterisitcs based estimation of $\epsilon^{L,R}$ }
\label{sec:characteristicbasedestimation}
Several choices for estimation of $\epsilon^{L,R}$ can be designed. Since the purpose of $\epsilon^{L,R}$ is primarily to increment the value of $\overline{S_L}$ and 
$\overline{S_R}$ locally in the vicinity of a shock wave and tend to zero elsewhere, any viable shock sensor is a good candidate for it.
Since shock instability is primarily a multidimensional phenomenon, it is also advisable to have an estimation of $\epsilon^{L,R}$ that takes this fact into account. 
This is motivated from the work of Sanders et al \cite{sanders1998} who have shown through linear analysis that for upwind schemes in multidimensional computations, 
a multidimensional shock sensor is more effective in ensuring shock stability. 

One possible estimate for $\epsilon^{L,R}$ can be based upon the idea of an entropy fix. Several authors \cite{peery1988,sanders1998,lin1995,pandolfi2001} have used such 
entropy fix based
estimates to save Roe scheme from shock instabilities.
Following \cite{pandolfi2001}, we define $\epsilon^{L,R}$ at an interface ($i,j+1/2$) as, 

\begin{align}
 \epsilon_{i,j+1/2}^{L,R} = max(\eta_1,\eta_2,\eta_3,\eta_4,\eta_5)
\end{align}
where each $\eta_k$ ($k$=1...5) corresponds to the dissipation quantity calculated at predefined interfaces that form a stencil around interface ($i,j+1/2$) as shown 
in Fig.(\ref{fig:stencilforepsilon}). The stencil is chosen such that modified diffusion vector $\mathbf{\overline{D}}$ is automatically employed on all interfaces in the 
vicinity of the shock.
Each $\eta_k$ can be 
evaluated using a common entropy fix formula,
\begin{align}
 \eta_k = \frac{1}{2}max( |\lambda_p(\mathbf{U}_R) - \lambda_p(\mathbf{U}_L)| )
\end{align}
where $p$ iterates over all the characteristic wavespeeds $\lambda_p$ of the Riemann problem at $k^{th}$ interface. An $\epsilon^{L,R}$ at any interface defined 
in the above form represents the largest one dimensional entropy correction of all the associated interfaces that forms a predefined stencil around it. Also this definition
ensures that an equal amount of dissipation is provided to both $S_L$ and $S_R$.
Note that the formula defined here is local solution dependant and hence is capable of admitting local variations in the flow field.
Futher the dissipation provided through $\epsilon^{L,R}$ has a multidimensional flavor because the dissipation introduced at the longitudinal interfaces are a function of the 
solution variation in the transverse direction and vice versa \cite{sanders1998}. The HLLEM-SWM scheme that uses this kind of $\epsilon^{L,R}$ will hereby be addressed as HLLEM-SWM-E 
(E stands for Eigenvalue based). It is important to emphasize here that in contrast to methods that target the eigenvalues of the linear wavefields to increase the 
overall dissipation of the 
scheme as in \cite{sanders1998,pandolfi2001,park2003}, the technique used here introduces dissipation through the numerical estimates of the nonlinear fields. 
Although the former strategy works well, there exist no physical justification for doing so. However the strategy presented here can be physically 
justified by the fact that in the two wave formulation of the underlying HLL scheme, $S_L$ and $S_R$ are the numerical 
characteristics associated with the shock waves and hence any dissipation infused through them would directly affect shock capturing ability.

	    \begin{figure}[ht]
	    \centering
	    \includegraphics[scale=0.4]{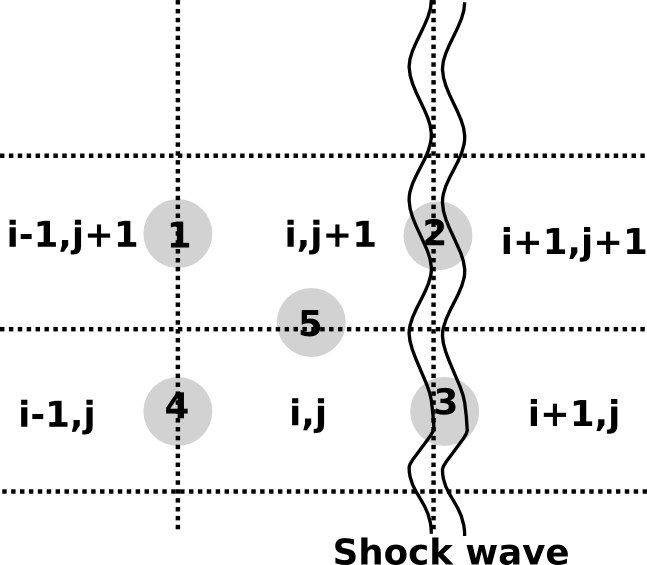}
	   \caption{Stencil adopted for calculating $\eta$'s required for estimation of $\epsilon^{L,R}$ at ($i,j+1/2$) interface.}
	   \label{fig:stencilforepsilon}
	    \end{figure}


\subsubsection{Pressure based estimation of $\epsilon^{L,R}$}
\label{sec:pressurebasedestimation}

Although quite appealing, the entropy fix based estimation of $\epsilon^{L,R}$ may cause loss of accuray in shear dominated viscous flows where the value of $\epsilon^{L,R}$, 
instead of tending 
towards zero within the shear layers, could become a non-zero quantity thereby introducing erroneous dissipation into the scheme and compromising on the accuracy of 
the solution. An interesting alternative to this would be to define $\epsilon^{L,R}$ based on pressure differences instead of eigenvalue differences. Hence we define 
$\epsilon^{L,R}$ through a pressure based shock sensor as,

\begin{align}
 \epsilon_{i,j+1/2}^{L,R} = (1-\omega_{i,j+1/2})|S_{L,R}|
 \label{eqn:epsilonomegarelation}
\end{align}
where $\omega_{i,j+1/2}$ is a pressure based shock sensor that can be constructed as \cite{zhang2017},
\begin{align}
 \omega_{i,j+1/2} = min_k (f_k),\ \ \ k=1...5
 \label{eqn:definitionofomega}
\end{align}
where $f_k$'s are pressure ratio based functions evaluated on a predefined stencil around $({i,j+1/2})$ interface shown in Fig.(\ref{fig:stencilforepsilon}). 
At $k^{th}$ interface, $f_k$ is defined as,  
\begin{align}
 f_k = min\left ( \frac{p_R}{p_L},\frac{p_L}{p_R} \right )_k^\beta
 \label{eqn:definitionofpressureratiofucntion}
\end{align}
Here $p_R$ and $p_L$ denotes the right and left cell center pressures across $k^{th}$ interface. A value of $\beta=5.0$ is used as per suggestions in
\cite{sangeeth2018_HLLCADC}. 
Note that this pressure based $\epsilon^{L,R}$ also retains the merits of the shock sensor discussed in 
Sec.(\ref{sec:characteristicbasedestimation}) ie, a multidimensional stencil
and local solution dependence. Also in contrast to characteristics based estimation discussed in Sec.(\ref{sec:characteristicbasedestimation}), here $S_L$ and $S_R$ are modified by unequal distribution of 
dissipation. This quantity of dissipation is designed to be proportional to the respective largest ($S_R$) and smallest ($S_L$) wave speed estimates at each interface. 
However, across typical linear wave fields without pressure jumps like contact and shear waves, this sensor would ideally ensure no loss of 
accuracy. An HLLEM-SWM scheme that uses this kind of $\epsilon^{L,R}$ will hereby be addressed as HLLEM-SWM-P (where P stands for Pressure based).
\sloppy

\begin{figure}[H]
	    \setcounter{subfigure}{0}
	    \subfloat[ $\mathbf{HLLEM-SWM}$ $\boldsymbol{\left(\hat{\rho} = 0.01, \hat{u} = 0, \hat{p} = 0\right)} $ ]{\label{fig:HLLEMSWM_density_perturbation}\includegraphics[scale=0.30]{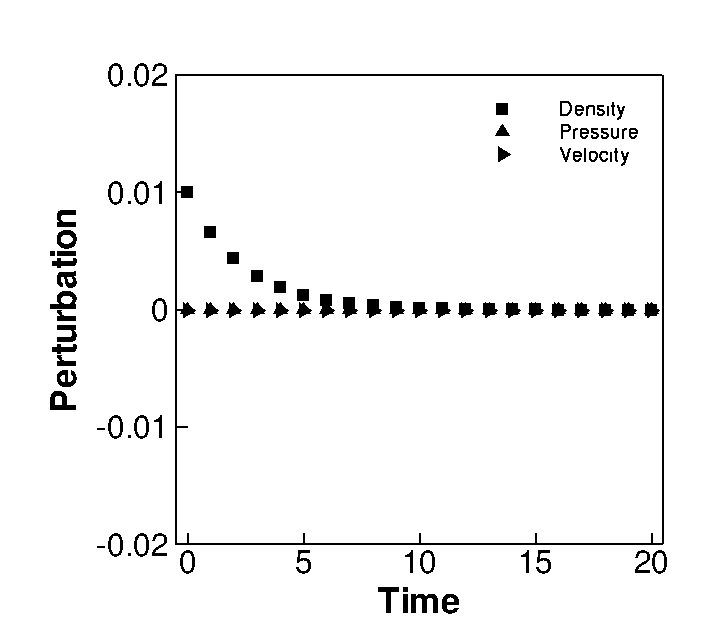}}
            \qquad
            \qquad
	    \subfloat[$\mathbf{HLLEM-SWM}$ $\boldsymbol{\left( \hat{\rho} = 0, \hat{u} = 0.01, \hat{p} = 0 \right) }$]{\label{fig:HLLEMSWM_velocity_perturbation}\includegraphics[scale=0.30]{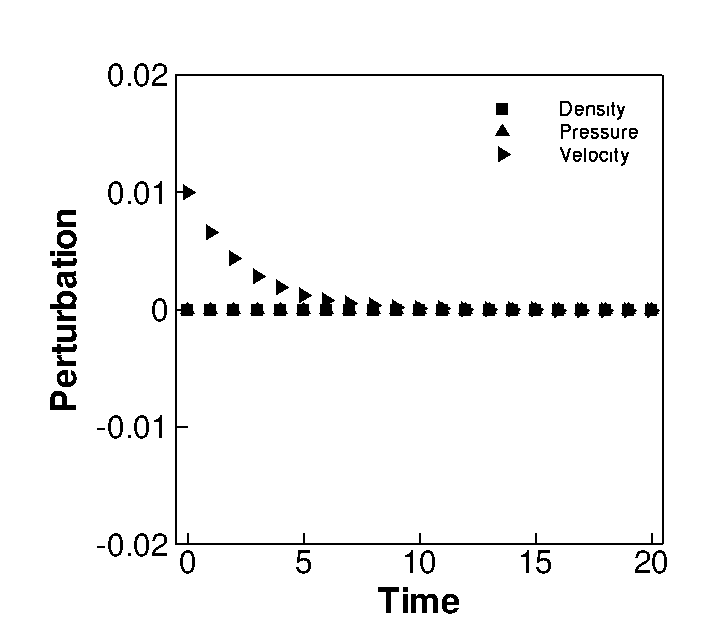}}\\
            \subfloat[$\mathbf{HLLEM-SWM}$ $\boldsymbol{ \left(\hat{\rho} = 0, \hat{u} = 0, \hat{p} = 0.01\right) }$]{\label{fig:HLLEMSWM_pressure_perturbation}\includegraphics[scale=0.30]{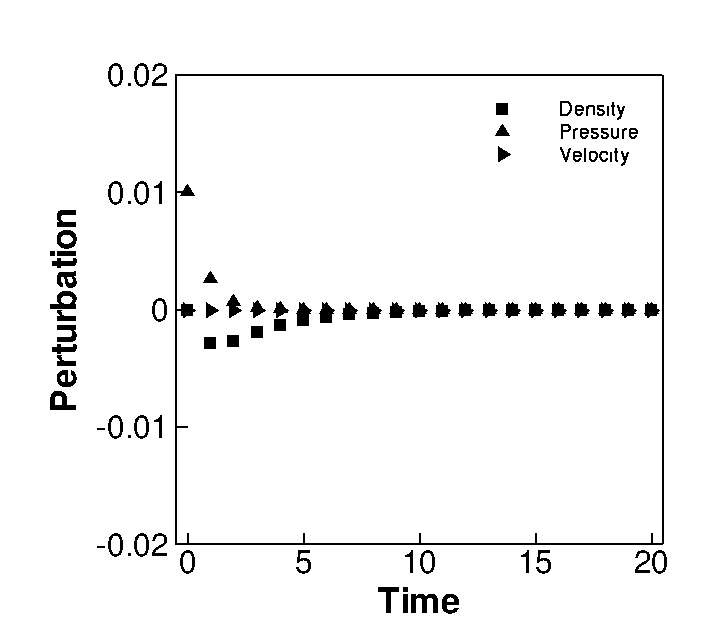}} 
	    \caption{Comparison of evolution of density, x-velocity and pressure perturbations in HLLEM-SWM class of schemes.}
	    \label{fig:perturbationstudies_HLLEMSWM}
	    \end{figure}

The perturbation evolution equations for HLLEM-SWM class of schemes are given in Table(\ref{table:evolutionequations}). 
It is interesting to note that all the three evolution equations of HLLEM-SWM scheme essentially comprise of HLLEM component and extra terms that contain 
the parameters $\epsilon^{L,R}$ and $\alpha$. The amplification factors for the perturbations $\hat{\rho},\hat{u},\hat{p}$ are respectively $\left( 1- \frac{2\lambda\alpha\epsilon^{L,R}}{\sqrt{\gamma}},
1-\frac{2\lambda\alpha\epsilon^{L,R}}{\sqrt{\gamma}}, 1-2\lambda(1+\frac{\alpha\epsilon^{L,R}}{\sqrt{\gamma}}) \right)$. From these, a strict von-Neumann like stablility criterion on $\lambda$ can 
be derived as,
\begin{align}
  0\leq\lambda \leq\frac{1}{1+\frac{\alpha \epsilon^{L,R}}{\sqrt{\gamma}}}
  \label{eqn:stability_hllemswm}
\end{align}
For a $\lambda$ in these bounds, it can be seen that any value for $\epsilon^{L,R},\alpha >0$ will introduce a damping factor of 
$\frac{2\lambda\alpha\epsilon^{L,R}}{\sqrt{\gamma}}$ into each of these perturbations. Notice that while this additional factor damps $\hat{u}$, it provides 
an ancillary damping to $\hat{p}$ over the preexisting damping coefficient $(1-2\lambda)$. In case of $\hat{\rho}$, the additional damping factor 
not only ensures that independent perturbations in density are attenuated, but also that any residual perturbations 
arising in $\hat{\rho}$ due to action of $\hat{p}$ are also suppressed. Thus in general, the additional factor ensures that the perturbations in all the three 
primitive quantities die down in time. 
These behaviours can be clearly observed in Fig.(\ref{fig:perturbationstudies_HLLEMSWM}). The plots correspond to $\lambda = 0.2$,
$\alpha=3.5$ and $\epsilon^{L,R} =1.0$. Given these damping characterisitcs, 
the HLLEM-SWM class of schemes are very unlikely to produce variation of conserved quantity $\rho u$ in the vicinity of the shock profile. This would ensure that numerical solutions of 
HLLEM-SWM schemes would remain shock instability free.

\subsection{Shock stable HLLEM scheme based on antidiffusion control}
\label{sec:HLLEM-ADC_formulation}
Since antidiffusion terms in the mass and interface-normal momentum flux discretizations in the transverse direction of the shock front are seen to be instigating the 
instability, a direct strategy to achieve shock stability in the HLLEM scheme would be to simply
reduce the quantity of this term on these crucial transverse interfaces. 
This can be readily achieved by 
employing a shock sensor to switch off the antidiffusive terms near a shock. We can define a modified HLLEM scheme, hereby termed as HLLEM-ADC (where ADC stands for 
\textbf{A}nti \textbf{D}iffusion \textbf{C}ontrol), to be 
applied on mass and interface-normal momentum flux discretizations on these critical interfaces in the vicinity of a shock wave as,

\begin{align}
   \mathbf{F}_{HLLEM-ADC} =
\begin{cases}
   \frac{1}{2}(\mathbf{F}_L + \mathbf{F}_R) + \mathbf{D} + \omega \mathbf{A}, & \text{if } mass,\;interface-normal\;momentum\\
   \frac{1}{2}(\mathbf{F}_L + \mathbf{F}_R) + \mathbf{D} +  \mathbf{A}, & \text{if } energy,\;interface-tangent\;momentum
\end{cases}
  \label{eqn:hllc-modifiedform}
\end{align}
Our experience shows that unlike the cure
discussed in Section.(\ref{sec:HLLE_SWM_formulation}) which had to be applied on all the interfaces in the vicinity of the numerical shock to be effective, 
the cure suggested in this section need to be applied only on the interfaces that are transverse to the shock. This may be because we are directly targeting the major 
trigger for the instability by controlling the antidiffusive terms in these critical flux components. Hence propensity to develop an unphysical variation in conserved quantity $\rho u$
along the shock front itself is reduced drastically. To ensure that the antidiffusive terms are withdrawn automatically only on these interfaces, we redefine the shock sensor $\omega$ as,
\begin{align}
 \omega_{i,j+1/2} = min_k (f_k),\ \ \ k=1...4
 \label{eqn:definitionofomega_redefined}
\end{align}

	    \begin{figure}[ht]
	    \centering
	    \includegraphics[scale=0.4]{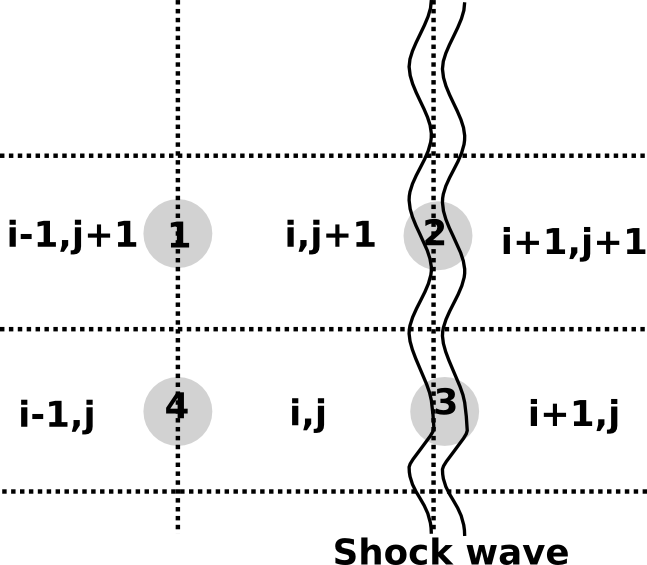}
	   \caption{Stencil adopted for calculating $\omega_{i,j+1/2}$ at interface $(i,j+1/2)$ for HLLC-ADC scheme.}
	   \label{fig:stencilforepsilon}
	    \end{figure}
\noindent Note that in the above formulation, the antidiffusion terms concerning both contact and shear 
waves are treated. However this is not the only possibility. Xie et al \cite{xie2017} have argued that within the antidiffusion term of the HLLEM scheme, 
it is the one concerning 
the shear wave resolution represented by $\delta_3 \tilde{\alpha}_3\tilde{R}_3$ that plays the major role in causing the onset of instability. 
Xie et al \cite{xie2017} use $\omega$ defined in Eq.(\ref{eqn:definitionofomega}) with $\alpha$ chosen to be 3.0 to control only the shear terms in all equations and on all 
interfaces in the vicinity of the shock. The resulting scheme is called HLLEMS by them.

The perturbation evolution equations for HLLEM-ADC scheme is given in Table(\ref{table:evolutionequations}). The evolution equations of HLLE-ADC in 
$\hat{\rho}$ and $\hat{u}$ turns out
to be those of the HLLE scheme with the additional antidissipation terms acted upon by the shock sensor $\omega$. The evolution of $\hat{p}$ resembles that of the HLLE and 
the HLLEM scheme.
The amplification factors for the perturbations 
$\hat{\rho},\hat{u},\hat{p}$ are respectively $\left( 1- 2\lambda(1-\omega),1-2\lambda(1-\omega),1-2\lambda \right)$. From these, a strict von-Neumann like stablility criterion on $\lambda$ can 
be derived as,
\begin{align}
  0\leq\lambda\leq\frac{1}{1-\omega}
  \label{eqn:stability_hllemadc}
\end{align}
Thus for $0\leq\omega<1$ and an appropriate choice of $\lambda$ under the above constraint, the perturbations 
$\hat{\rho}$ and $\hat{u}$ are acted upon by additional damping terms controlled by $\omega$. The perturbation $\hat{p}$ in itself is damped independent of the action of 
$\omega$. However, the rate of $\hat{p}$ feeding into $\hat{\rho}$ is minimized by $\omega$ leaving no scope for residual $\hat{\rho}$ to develop as a result of undamped
$\hat{p}$ in the flow field. When $\omega$ is equal to 0, we recover the HLLE behaviour with the 
antidiffusive behaviour completely suppressed. 

\begin{figure}[H]
	    \setcounter{subfigure}{0}
	    \subfloat[ $\mathbf{HLLEM-ADC}$ $\boldsymbol{\left(\hat{\rho} = 0.01, \hat{u} = 0, \hat{p} = 0\right)} $ ]{\label{fig:HLLEM_ADC_density_perturbation}\includegraphics[scale=0.30]{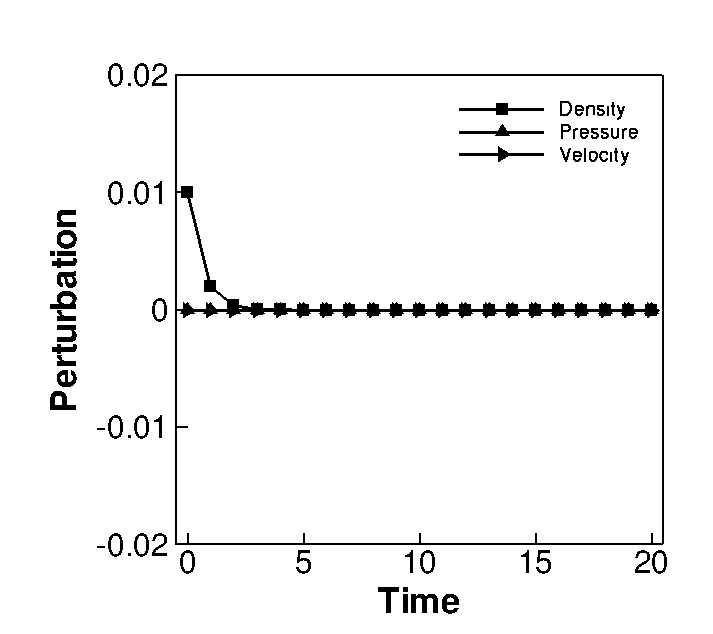}}
            \qquad
            \qquad
            \subfloat[$\mathbf{HLLEMS}$ $ \boldsymbol{\left(\hat{\rho} = 0.01, \hat{u} = 0, \hat{p} = 0\right) } $]{\label{fig:HLLEMS_density_perturbation}\includegraphics[scale=0.30]{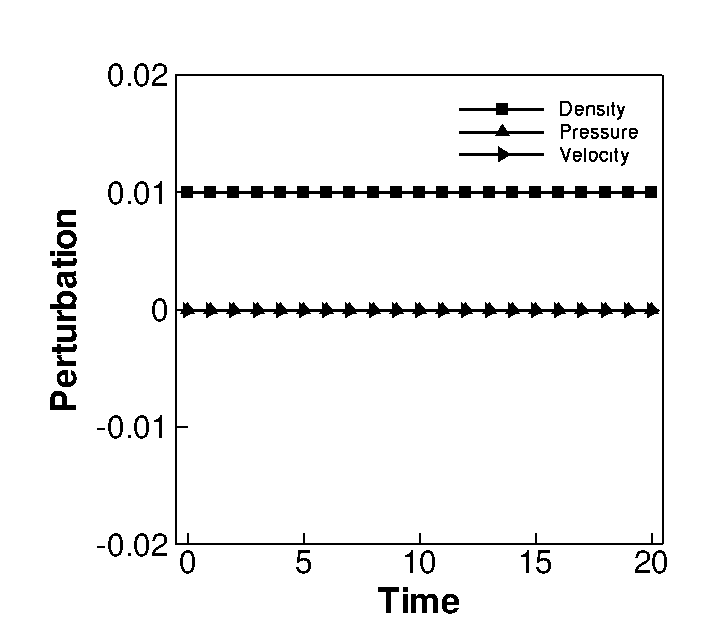}}\\
            \subfloat[$\mathbf{HLLEM-ADC}$ $\boldsymbol{\left( \hat{\rho} = 0, \hat{u} = 0.01, \hat{p} = 0 \right) }$]{\label{fig:HLLEM_ADC_velocity_perturbation}\includegraphics[scale=0.30]{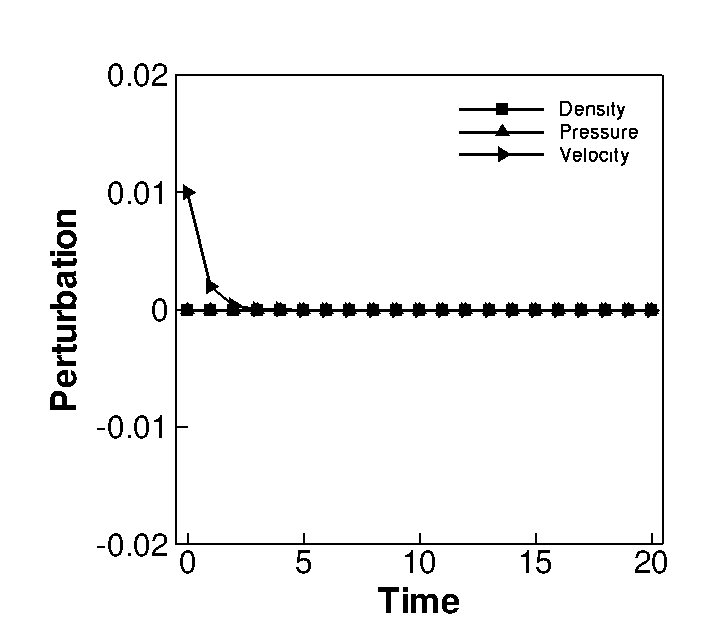}}
            \qquad
            \qquad
            \subfloat[$\mathbf{HLLEMS}$  $\boldsymbol{\left(\hat{\rho} = 0, \hat{u} = 0.01, \hat{p} = 0\right) }$]{\label{fig:HLLEMS_velocity_perturbation}\includegraphics[scale=0.30]{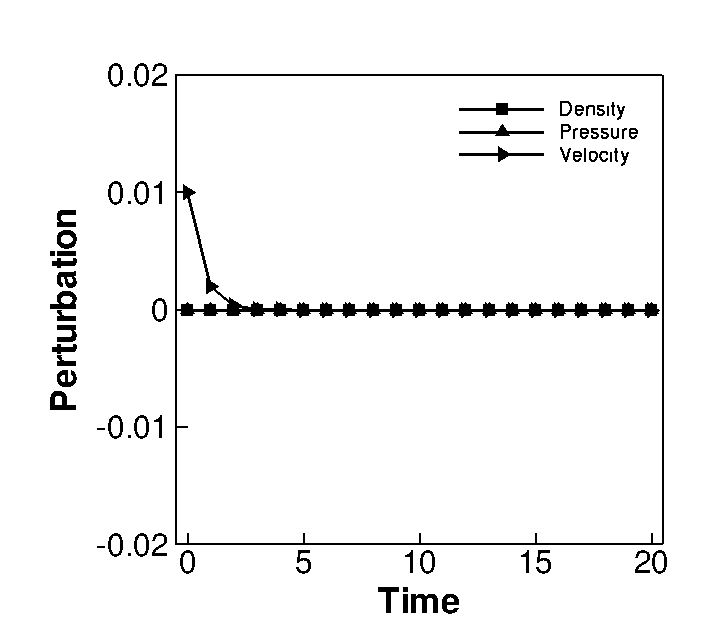}}\\
            \subfloat[$\mathbf{HLLEM-ADC}$ $\boldsymbol{ \left(\hat{\rho} = 0, \hat{u} = 0, \hat{p} = 0.01\right) }$]{\label{fig:HLLEM_ADC_pressure_perturbation}\includegraphics[scale=0.30]{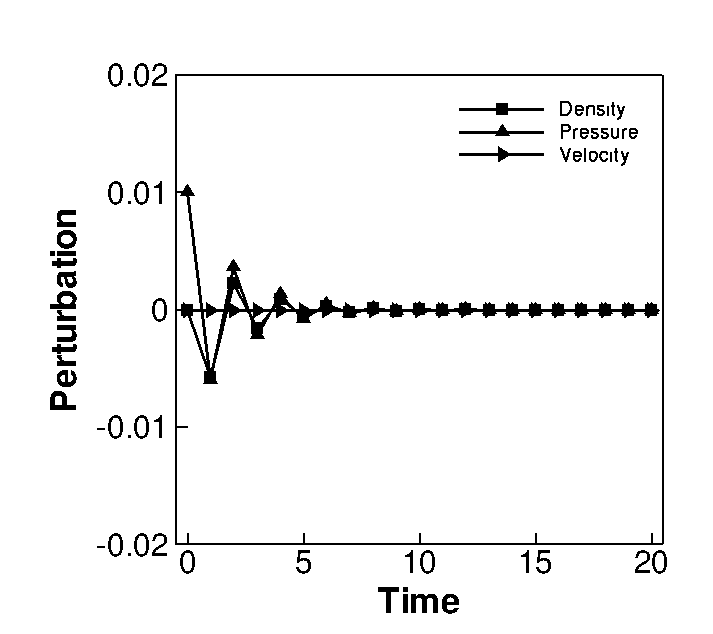}} 
	    \qquad
            \qquad
            \subfloat[$\mathbf{HLLEMS}$ $\boldsymbol{ \left(\hat{\rho} = 0, \hat{u} = 0, \hat{p} = 0.01\right) } $]{\label{fig:HLLEMS_pressure_perturbation}\includegraphics[scale=0.30]{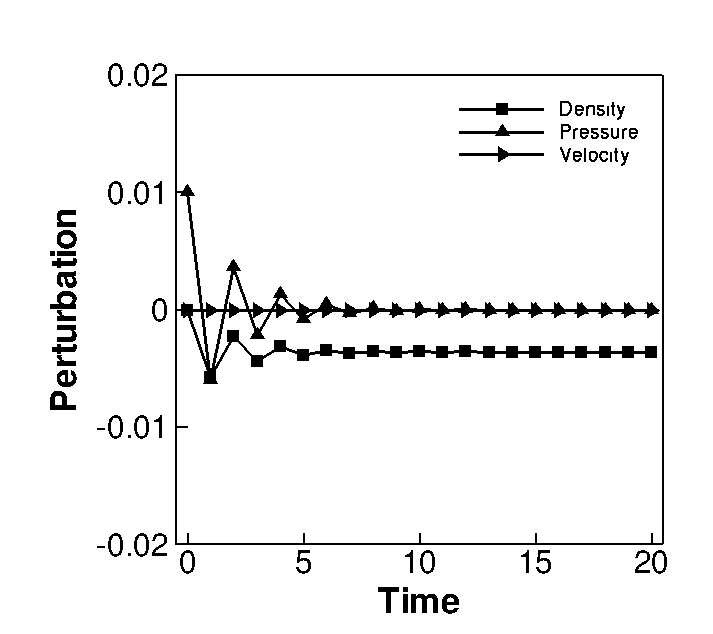}}
	    \caption{Comparison of evolution of density, x-velocity and pressure perturbations in HLLEM-ADC and HLLEMS schemes.}
	    \label{fig:perturbationstudies_hllem_adc+hllems}
	    \end{figure}

We also present the evolution equations of HLLEMS scheme in Table \ref{table:evolutionequations} which was not derived in \cite{xie2017}. 
The corresponding  amplification factors for the perturbations 
$\hat{\rho},\hat{u},\hat{p}$ are respectively $\left( 1,1-2\lambda(1-\omega),1-2\lambda \right)$.
It is seen that in comparision to HLLEM-ADC scheme, the HLLEMS scheme allows the action of $\omega$ only on $\hat{u}$ perturbation and not on $\hat{\rho}$.
The difference between HLLEM-ADC and HLLEMS can be clearly seen from the behavior of $\hat{\rho}$. While 
HLLEM-ADC scheme is able to employ $\omega$ to damp the perturbation $\hat{\rho}$, HLLEMS does not have such provisions. Hence any $\hat{\rho}$ in the solution that is generated
will persist in time. Further, although $\hat{p}$ is damped similar to HLLEM-ADC scheme, its effect on $\hat{\rho}$ is not suppressed.  
Thus in overall, density perturbations, those that are generated independently 
and those induced by the feeding of $\hat{p}$, both remain undamped in case of the HLLEMS scheme. The difference in the evolution equations of these schemes can be more clearly noticed in Fig.(\ref{fig:perturbationstudies_hllem_adc+hllems}). 
As reported in \cite{sangeeth2018_HLLCADC} the inability to damp density perturbations could then theoretically cause variations in conserved quantity $\rho u$ near shocks and 
eventually lead to instability. 
The difference between the quality of solutions produced by these schemes will become evident in certain numerical examples shown later in Section.(\ref{sec:numericaltests}).

\section{Numerical results}
\label{sec:numericaltests}
In this section, we demonstrate the efficacy of the proposed HLLC-SWM and HLLC-ADC schemes through a series of strict shock instability test problems on which many 
contact-shear preserving Riemann solvers fail. Along with the original HLLEM scheme, solutions for HLLEMS scheme\cite{xie2017} are also provided for comparison in all cases. It has been reported in \cite{gressier2000} that
the intensity of instability manifestation in certain test problems reduces as the order of accuracy of the computation increases; although it does not completely remove it. Based on this,
certain test cases are computed as plain first order while second order solutions are sought for others. Second order spatial accuracy is achieved by limiting gradients 
of primitive variables obtained using
Green Gauss method \cite{balzek2005} with Barth Jersperson limiter \cite{barth1989}. Second order time accuracy is achieved using 
strong stability preserving variant of Runge Kutta method \cite{Gottlieb2001}. All boundary conditions are set using ghost cells.

\subsection{Moving shock instability problem}
\label{sec:movingshock_hllemfixed}
\begin{figure}[H]
	    \centering
	    \setcounter{subfigure}{0}
	    \subfloat[HLLEM]{\label{fig:hllem_movingshock}\includegraphics[scale=0.27]{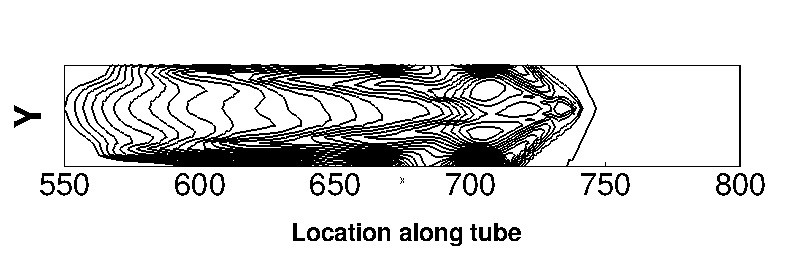}}
	    \subfloat[HLLEM-SWM-P]{\label{fig:hllem_swm_p_movingshock}\includegraphics[scale=0.27]{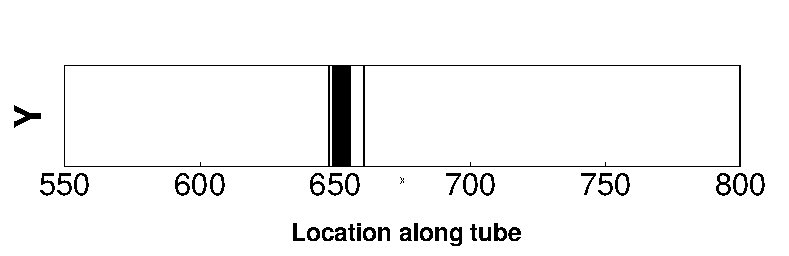}}\\
	    \subfloat[HLLEM-SWM-E]{\label{fig:hllem_swm_e_movingshock}\includegraphics[scale=0.27]{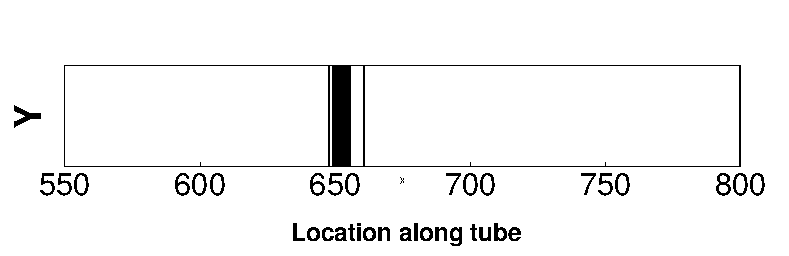}}
	    \subfloat[HLLEM-ADC]{\label{fig:hllem_adc_movingshock}\includegraphics[scale=0.27]{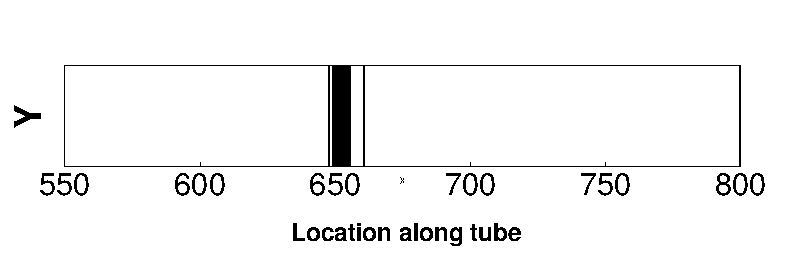}}\\
	    \subfloat[HLLEMS]{\label{fig:hllems_movingshock}\includegraphics[scale=0.27]{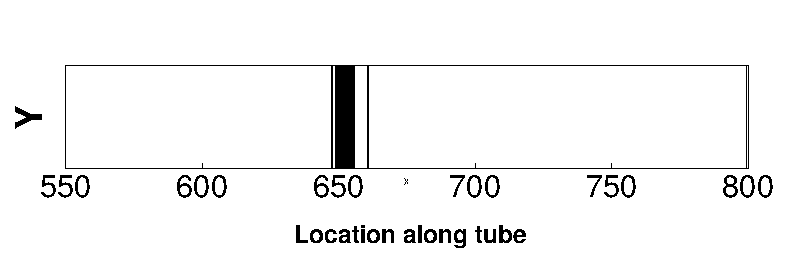}}
	   \caption{Density contours for M=6 Quirk's odd-even decoupling problem.}
	   \label{fig:movingshockresults}
	    \end{figure}

Shock instability that occurs in a moving shock propagating down a computational tube was first reported by Quirk \cite{quirk1994}. In the present
case the strength of the shock is chosen to be M=6 and the computational domain consists of 800 by 20 cells in x and y directions respectively.
The shock is made to propagate into a stationary fluid with physical state $(\rho,u,v,p)=(1.4,1.0,0.0,1.0)$.
The instability is triggered by perturbing the centerline grid of the computational domain to an order of 1E-6. First order solution is sought. 
The CFL for the computations were
taken to be 0.5 and simulations were run till shock reached a location x=650. The results showing fifty density
contours equally spanning values from 1.4 to 7.34 is shown in Fig.(\ref{fig:movingshockresults}).

\noindent Fig.(\ref{fig:hllem_movingshock}) clearly shows the deteriorated condition of the shock profile calculated by the HLLEM scheme. 
Fig.(\ref{fig:hllem_swm_p_movingshock}) and Fig.(\ref{fig:hllem_swm_e_movingshock}) demonstrates that both versions of HLLC-SWM scheme are free
from instability on this problem. Fig.(\ref{fig:hllem_adc_movingshock}) and Fig.(\ref{fig:hllems_movingshock}) show that both full and partial control of antidissipation 
removes instability on this problem.

\subsection{Double Mach Reflection problem }
\label{sec:dmr_hllemfixed}
\begin{figure}[H]
	    \centering
	    \setcounter{subfigure}{0}
	    \subfloat[HLLEM]{\label{fig:hllem_dmr}\includegraphics[scale=0.3]{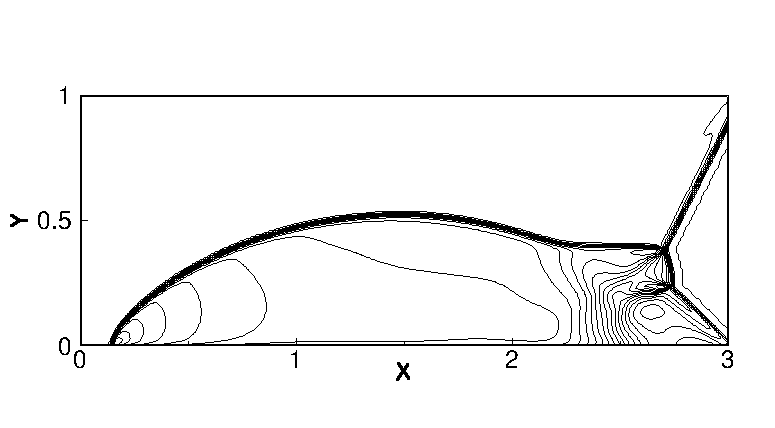}}
	    \subfloat[HLLEM-SWM-P]{\label{fig:hllem_swm_p_dmr}\includegraphics[scale=0.3]{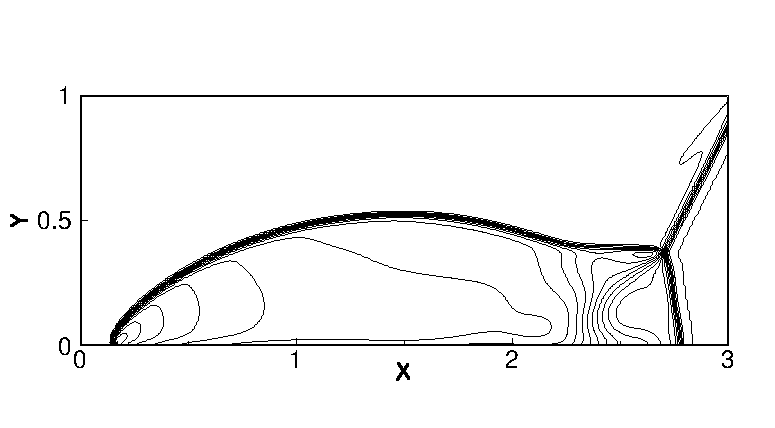}}\\
	    \subfloat[HLLEM-SWM-E]{\label{fig:hllem_swm_e_dmr}\includegraphics[scale=0.3]{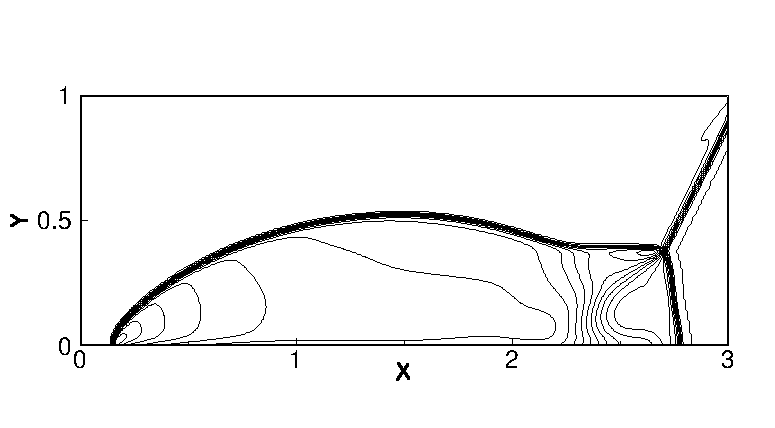}}
	    \subfloat[HLLEM-ADC]{\label{fig:hllem_adc_dmr}\includegraphics[scale=0.3]{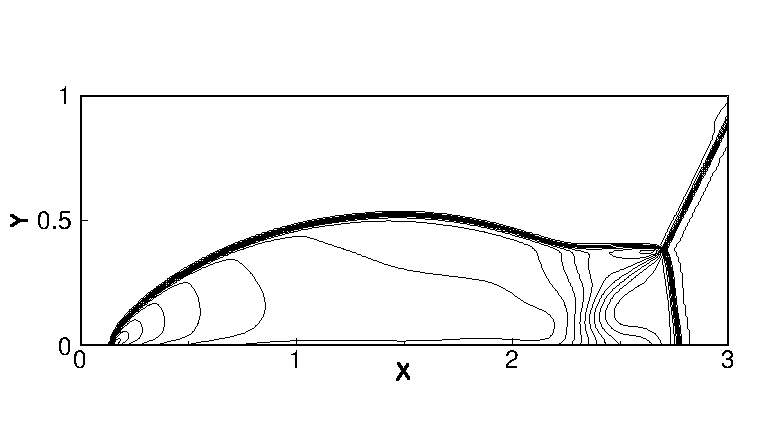}}\\
	    \subfloat[HLLEMS]{\label{fig:hllems_dmr}\includegraphics[scale=0.3]{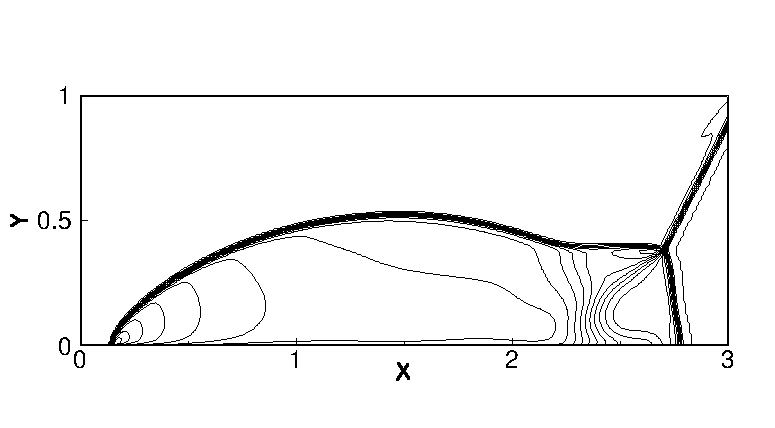}}
	   \caption{Density contours for double Mach reflection problem.}
	   \label{fig:dmrresults}
	    \end{figure}

To demonstrate the robustness of the proposed schemes, another standard test problem called Double Mach reflection problem is used \cite{woodward1984}.
For the present test, a domain of $4.0\times1.0$ is chosen and is constituted of $480\times120$ structured Cartesian cells. An 
oblique shock corresponding to $M=10$, making a $60^o$
angle with the bottom wall at $x=0.16667$ is made to propagate through the domain. Cells ahead of the shock are initialized with values 
$(\rho,u,v,p) = (1.4,0,0,1)$ while those after the shock are set to appropriate post shock conditions. Inlet boundary is maintained at post shock 
conditions while zero gradient condition is used at outlet boundary. Top boundary conditions are adjusted to allow for propagation of shock front. 
At the bottom, post shock conditions are maintained till $x=0.16667$ after which invisid wall conditions are used. The simulation is run till 
$t=0.02$ with CFL of 0.8. The problem is computed to first order accuracy. Fig.(\ref{fig:dmrresults}) shows results of the experiment with twenty five density contours equally spanning values 
from 1.4 to 21.
Fig.(\ref{fig:hllem_dmr}) clearly shows the presence of kinked Mach stem and the subsequent triple point produced by HLLEM scheme. The solutions computed by 
HLLEM-SWM schemes, HLLC-ADC and HLLEMS show no trace of instability.

\subsection{ Hypersonic flow over a blunt body}
\label{sec:bluntbody_hllemfixed}
\begin{figure}[H]
	    \centering
	    \setcounter{subfigure}{0}
	    \subfloat[HLLEM]{\label{fig:hllem_bluntbody_centerline_perturb}\includegraphics[scale=0.29]{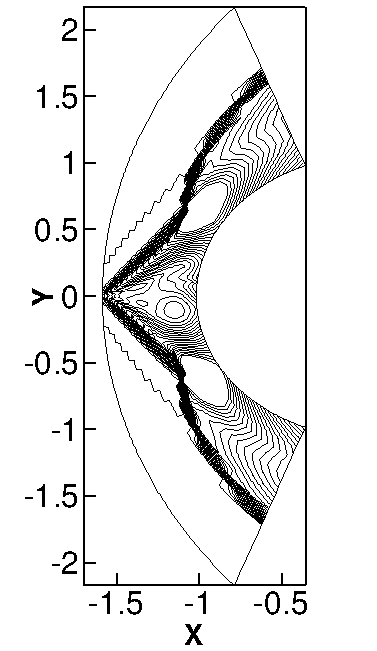}}
	    \subfloat[HLLEM-SWM-P]{\label{fig:hllem_swm_p_bluntbody_centerline_perturb}\includegraphics[scale=0.3]{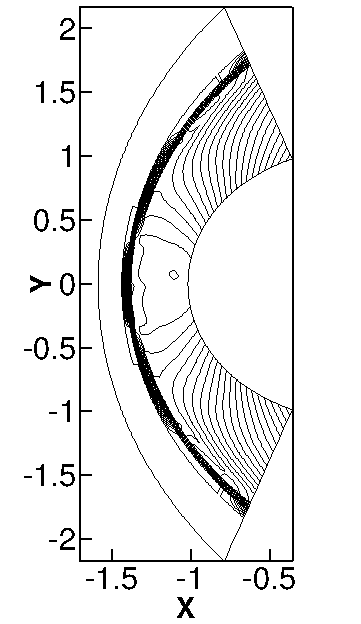}}
	    \subfloat[HLLEM-SWM-E]{\label{fig:hllem_swm_e_bluntbody_centerline_perturb}\includegraphics[scale=0.3]{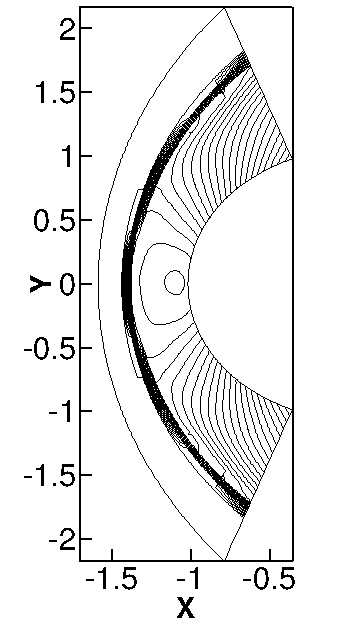}}\\
	    \subfloat[HLLEM-ADC]{\label{fig:hllem_adc_bluntbody_centerline_perturb}\includegraphics[scale=0.3]{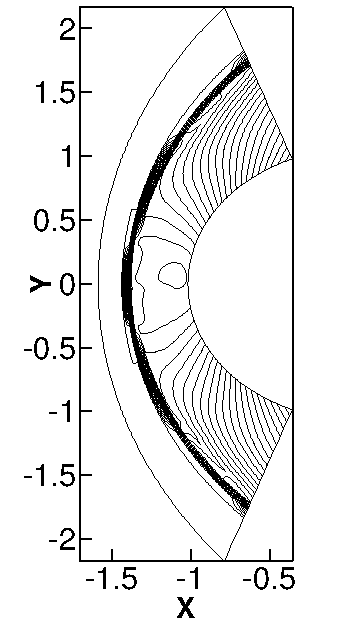}}
	    \subfloat[HLLEMS]{\label{fig:hllems_bluntbody_centerline_perturb}\includegraphics[scale=0.3]{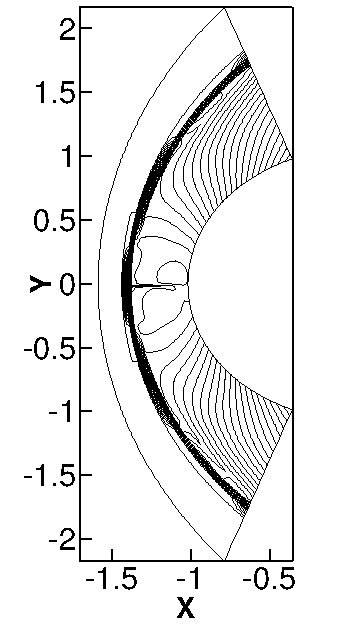}}
	   \caption{Density contours for M=20 supersonic flow over a cylindrical body.}
	   \label{fig:bluntbody_centerline_perturbresults}
	    \end{figure}

Another routine test problem that is used to investigate the susceptibility of a numerical scheme to shock instability is the steady state 
numerical solution of a supersonic or hypersonic flow over cylindrical body. The instability in this problem manifests as distorted contour lines and in certain
extreme cases, complete obliteration of the shock profile. The unstable soluton will result in erroneous stagnation values of physical quantities which may 
eventually affect heat transfer predictions \cite{kitamura2010}. 
The problem is set up by placing a cylindrical body of radius 1 unit in a M=20 flow with free stream conditions given as 
$(\rho,u,v,p) = (1.4,20.0,0.0,1.0)$.
The computational mesh is prepared using method described in \cite{huang2011} wherein 320$\times$40 body fitted structured quadrilateral cells 
are used in circumferential and radial directions respectively. To instigate the instabilities, we perturb the radial grid line that lies along y=0 in a saw-tooth 
profile at the order of 1E-4. The inlet boundary is maintained as supersonic inlet while at the solid wall, impermeability is prescribed
with density and pressure are extrapolated 
from the internal cell. Simple extrapolation
is employed at top and bottom boundaries. The computation of this problem is carried out to first order accuracy. The CFL for the computations were taken to be 0.5 
and simulations were run for 
20,000 iterations. The results showing twenty density contours equally spanning value from 1.4 to 8.5 is shown in Fig.(\ref{fig:bluntbody_centerline_perturbresults}). 
	
Fig.(\ref{fig:hllem_bluntbody_centerline_perturb}) clearly shows the occurrence of the Carbuncle phenomenon in the solution computed by the HLLEM scheme. Amongst all the schemes, 
solution obtained using the HLLC-SWM-E given in Fig.(\ref{fig:hllem_swm_e_bluntbody_centerline_perturb}) seems to be the most smoothest. HLLC-SWM-P also produces quite comparable
solution to the HLLC-SWM-E scheme. On the other hand, the difference between full and partial antidiffusion control strategies used by the HLLC-ADC and the HLLEMS schemes respectively
is evident on this test case. While the solution computed by HLLC-ADC in Fig.(\ref{fig:hllem_adc_bluntbody_centerline_perturb}) shows no signs of perturbations, 
the one computed by HLLEMS in Fig.(\ref{fig:hllems_bluntbody_centerline_perturb}) shows clear asymmetry near the centerline. This could indicate that 
switching off antidiffusion terms corresponding to the shear waves alone may not be sufficient for complete removal of instability.

\subsection{Diffraction of a moving normal shock over a 90$^{0}$ corner}
\label{sec:supersoniccorner_hllemfixed}
\begin{figure}[H]
	    \centering
	    \setcounter{subfigure}{0}
	    \subfloat[HLLEM]{\label{fig:hllem_supersoniccorner}\includegraphics[scale=0.3]{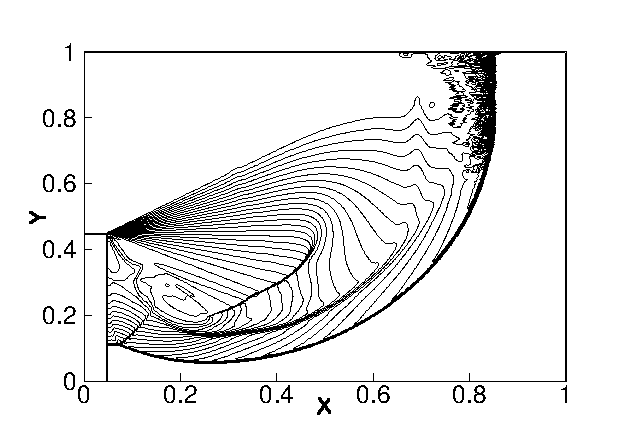}}
	    \subfloat[HLLEM-SWM-P]{\label{fig:hllem_swm_p_supersoniccorner}\includegraphics[scale=0.3]{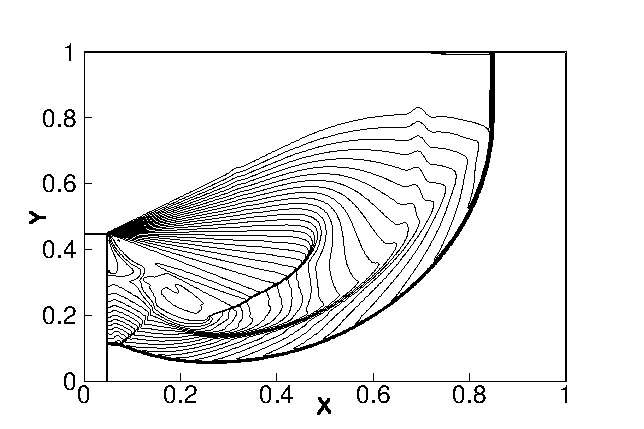}}\\
	    \subfloat[HLLEM-SWM-E]{\label{fig:hllem_swm_e_supersoniccorner}\includegraphics[scale=0.3]{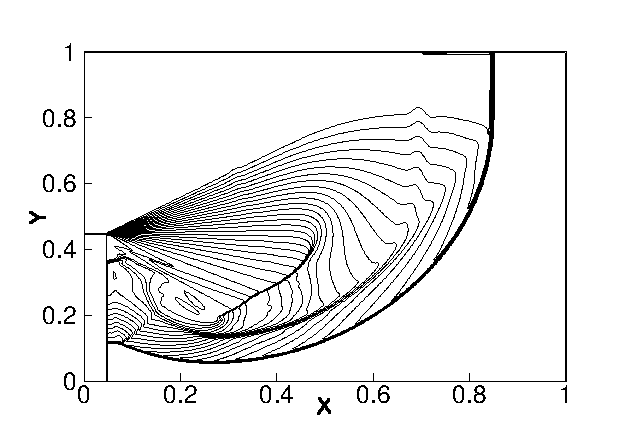}}
	    \subfloat[HLLEM-ADC]{\label{fig:hllem_adc_supersoniccorner}\includegraphics[scale=0.3]{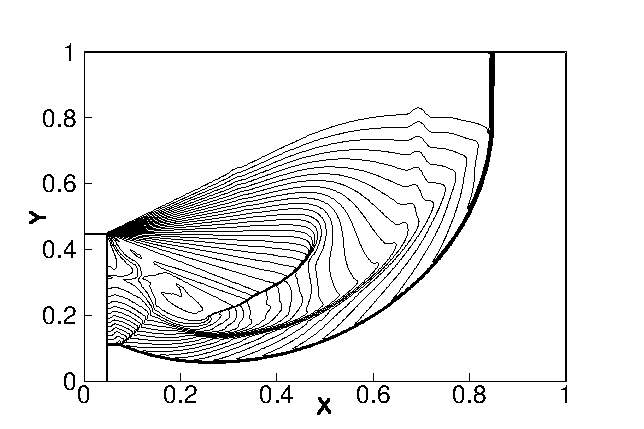}}\\
	    \subfloat[HLLEMS]{\label{fig:hllems_supersoniccorner}\includegraphics[scale=0.3]{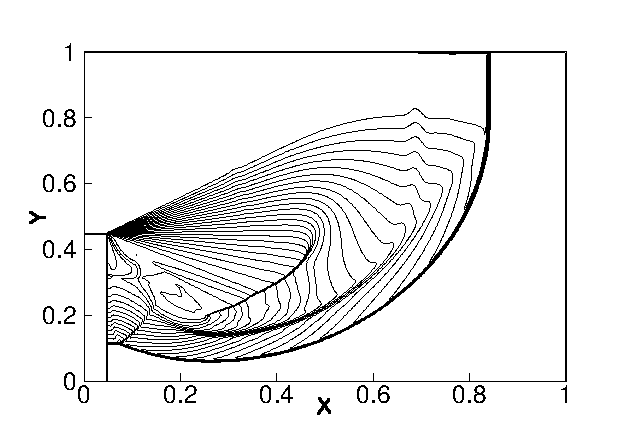}}
	   \caption{Density contours for M=5.09 normal shock diffraction around a $90^o$ corner.}
	   \label{fig:supersoniccornerresults}
	    \end{figure}

The problem of a Mach 5.09 normal shock's sudden expansion around a $90^o$ corner was studied in \cite{quirk1994}. A unit dimensional domain 
is meshed with 400 $\times$ 400 cells. The right angled corner is located at $x=0.05, y=0.45$. To begin with, the normal shock is located at $x=0.05$. Remaining 
cells are initialized to a stationary fluid with properties $\rho=1.4, u=0.0, v=0.0$ and $p=1.0$. The inlet boundary is maintained as post shock 
conditions while outlet boundary is set to zero gradient. Top boundary is adjusted to allow shock propagation. Bottom boundary behind the corner uses extrapolated values from within the domain. The corner 
surface is maintained as reflective wall. The problem is computed to second order accuracy. Simulations are run for $t=0.1561$ units with CFL of 
0.4. Thirty density contours equally spanning values of 0 to 7 are
shown in Fig.(\ref{fig:supersoniccornerresults}).
Fig.(\ref{fig:hllem_supersoniccorner}) clearly demonstrates the extent of instability present in the computed shock profile produced by HLLEM scheme. A part of the normal shock in the right 
corner is completely distorted in this case. Although solutions have quite improved, some benign amount of perturbation is observed close to the top boundary in solutions
computed by HLLC-SWM schemes and HLLEMS scheme as seen in Figs.(\ref{fig:hllem_swm_p_supersoniccorner},\ref{fig:hllem_swm_e_supersoniccorner} and \ref{fig:hllems_supersoniccorner}).
The solution computed by HLLC-ADC seems to be the most cleanest among all the schemes. 

\subsection{ Supersonic flow over forward facing step}
\label{sec:ffs_hllemfixed}
\begin{figure}[H]
	    \centering
	    \setcounter{subfigure}{0}
	    \subfloat[HLLEM]{\label{fig:hllem_forwardfacingstep}\includegraphics[scale=0.3]{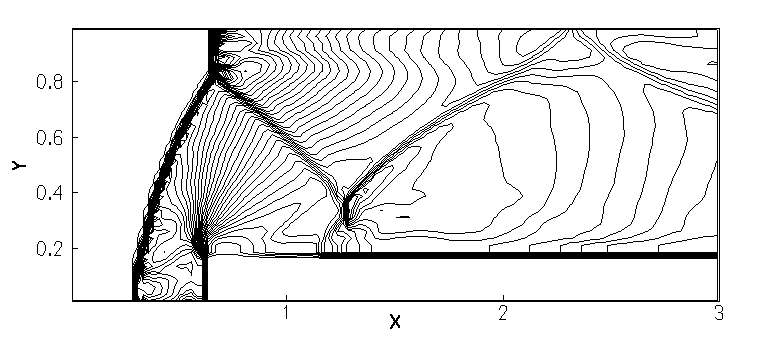}}
	    \subfloat[HLLEM-SWM-P]{\label{fig:hllem_swm_p_forwardfacingstep}\includegraphics[scale=0.3]{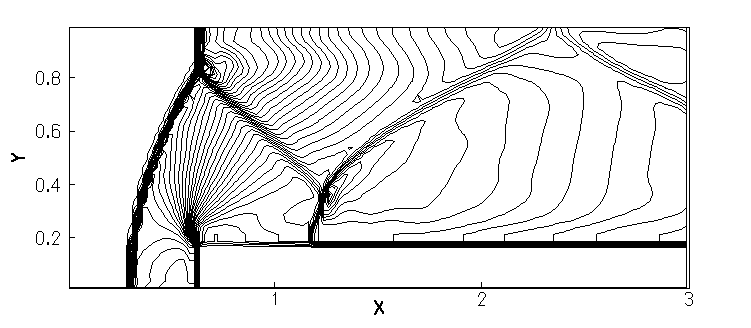}}\\
	    \subfloat[HLLEM-SWM-E]{\label{fig:hllem_swm_e_forwardfacingstep}\includegraphics[scale=0.3]{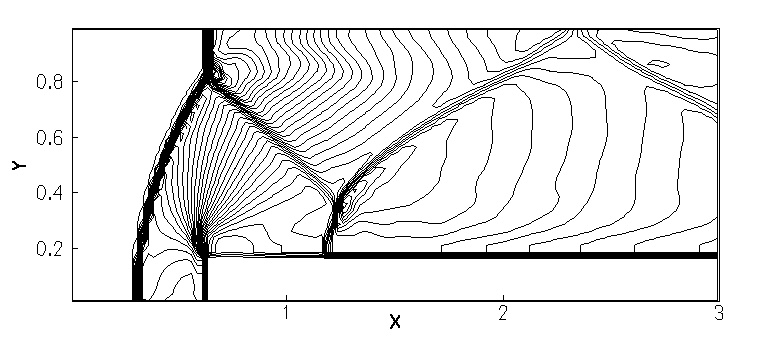}}
	    \subfloat[HLLEM-ADC]{\label{fig:hllem_adc_forwardfacingstep}\includegraphics[scale=0.3]{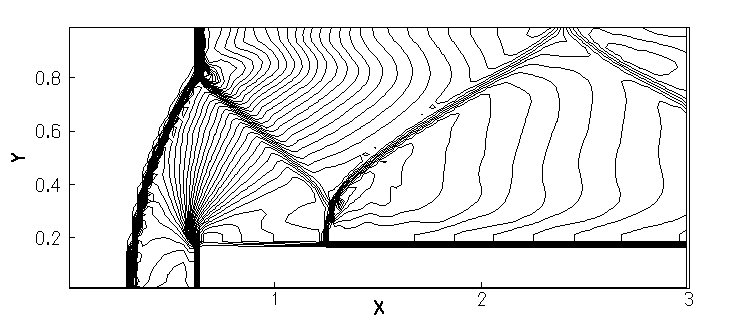}}\\
	    \subfloat[HLLEMS]{\label{fig:hllems_forwardfacingstep}\includegraphics[scale=0.3]{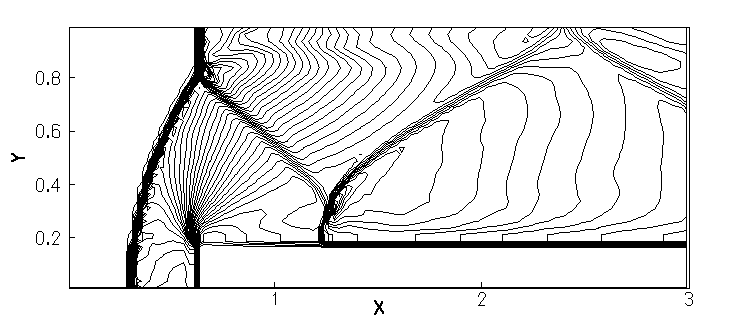}}
	   \caption{Density contours for M=5.09 normal shock diffraction around a $90^o$ corner.}
	   \label{fig:ffsresults}
	    \end{figure}

Supersonic flow over forward facing step is another standard problem that has been extensively studied earlier in 
\cite{woodward1984}. The computational domain is of size $3\times1$. A 0.2 units high step is located at a distance of 0.6 units from 
the inlet. The domain is meshed with $120\times40$ structured Cartesian cells. An initial value of $(\rho,u,v,p)= (1.4,3,0,1)$ that corresponds to a M=3 supersonic flow
is provided in the 
whole domain. The inlet boundary is maintained as freestream while the outlet boundary is set to zero gradient. The top and the bottom walls are
set as inviscid walls. The problem is computed to second order accuracy. The simulations are run for t=4 with CFL as 0.5. 
Fig.(\ref{fig:ffsresults}) shows forty equispaced density contours spanning 0.2 to 7.0. 
	    
It is seen from Fig.(\ref{fig:hllem_forwardfacingstep}) that predominant perturbations are visible in the shock profile computed by HLLEM scheme. These are most
prominent at the normal shock stems of the primary shock near the bottom and top boundaries. Further, HLLEM solution also has a severely distorted stem in the 
reflected shock. The solutions by HLLEM-SWM, HLLEM-ADC and HLLEMS shows marked improvement compared to that of HLLEM. In all the cases the perturbations in shock profile
is non existent and the reflected shock stem is captured well near the solid wall. 

\subsection{Inclined stationary shock instability problem}
\label{sec:inclinedshock_hllemfixed}
This problem was reported in \cite{ohwada2013}. In the present setup a stationary shock is initialized at an angle of $63.43^o$ with respect to the
global x-direction. The computation mesh consists of 50$\times$30 cells uniformly spanning a domain of size 50.0$\times$30.0.
The initial shock wave is set up along the line y=2(x-12). The conditions before the shock wave are provided using the conditions
$(\rho,u,v,p)_L=(1.0,447.26,-223.50,3644.31)$ and it describes a supersonic flow normal to the shock front. The after shock conditions are given as
$(\rho,u,v,p)_R=(5.444,82.15,-41.05,207725.94)$. The left and right boundaries are maintained as supersonic inlet and subsonic outlet respectively.
At the top boundary cells from 1 to 15 are maintained as supersonic inlet while all the remaining cells are set periodic to the corresponding 
bottom internal cells. At the bottom, cells from 36 to 50 are set as subsonic outlet while the remaining cells are set periodic to the corresponding 
cells at the top internal cells. CFL of the computation was chosen as 0.5 and first order solution was sought at t=5. 
Fig.(\ref{fig:inclinedshock}) shows the result of this experiment where thirty density contours 
uniformly spanning 1.0 to 5.44 are shown.

	  \begin{figure}[H]
	    \centering
	    \setcounter{subfigure}{0}
	    \subfloat[HLLEM]{\label{fig:hllem_inclinedshock}\includegraphics[scale=0.25]{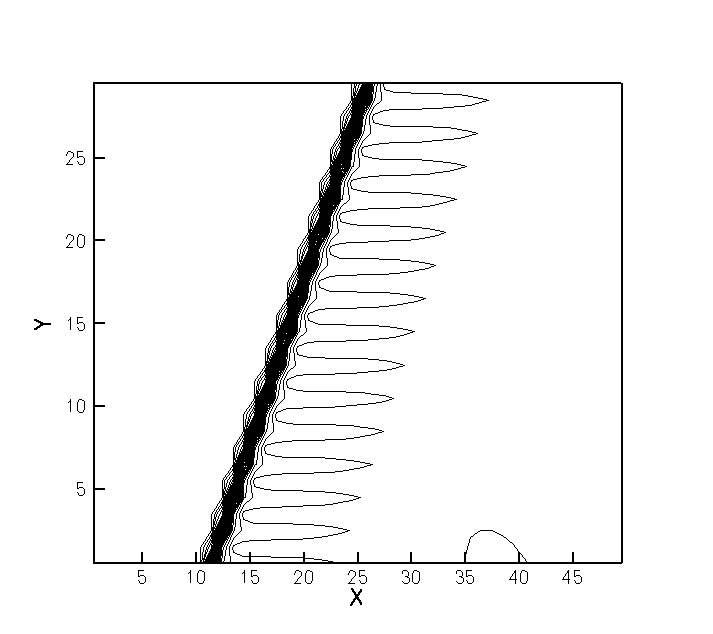}}
	    \subfloat[HLLEM-SWM-P]{\label{fig:hllem_swm_p_inclinedshock}\includegraphics[scale=0.25]{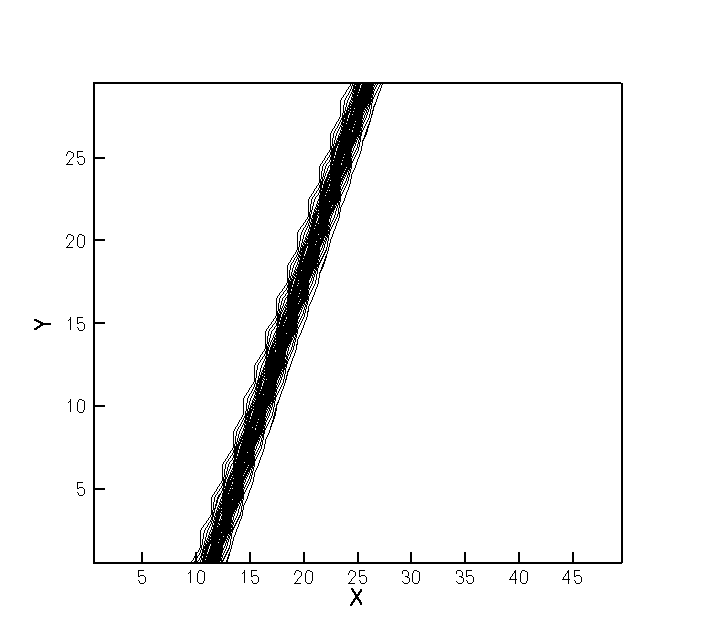}}\\
	    \subfloat[HLLEM-SWM-E]{\label{fig:hllem_swm_e_inclinedshock}\includegraphics[scale=0.25]{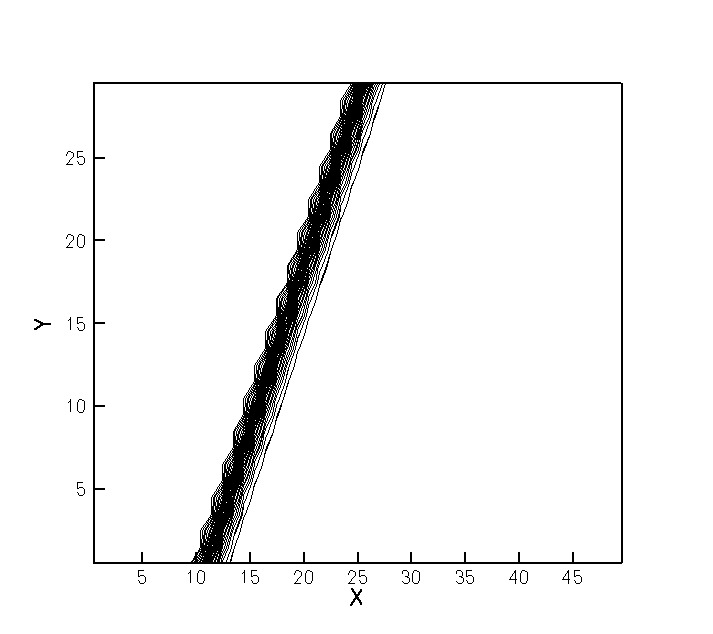}}
	    \subfloat[HLLEM-ADC]{\label{fig:hllem_adc_inclinedshock}\includegraphics[scale=0.25]{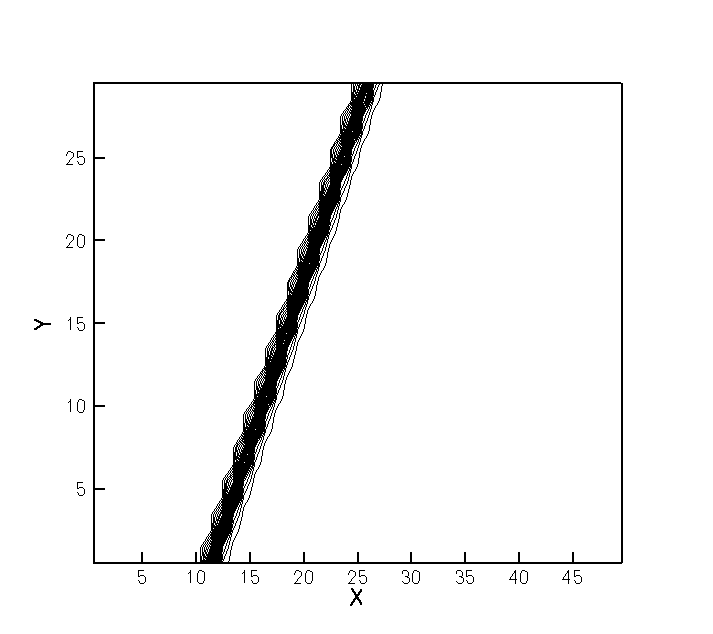}}\\
	    \subfloat[HLLEMS]{\label{fig:hllems_inclinedshock}\includegraphics[scale=0.25]{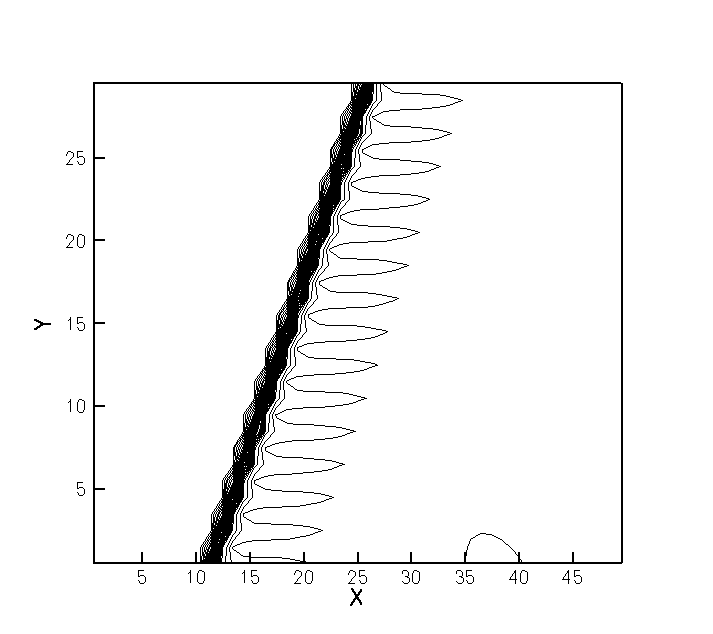}}
	   \caption{Density contours for M=7 stationary inclined shock problem.}
	   \label{fig:inclinedshock}
	    \end{figure}
	    
	    \begin{figure}[H]
	    \centering
	    \setcounter{subfigure}{0}
	    \subfloat[HLLEM]{\label{fig:hllem_inclinedshock_densityalongshock}\includegraphics[scale=0.3]{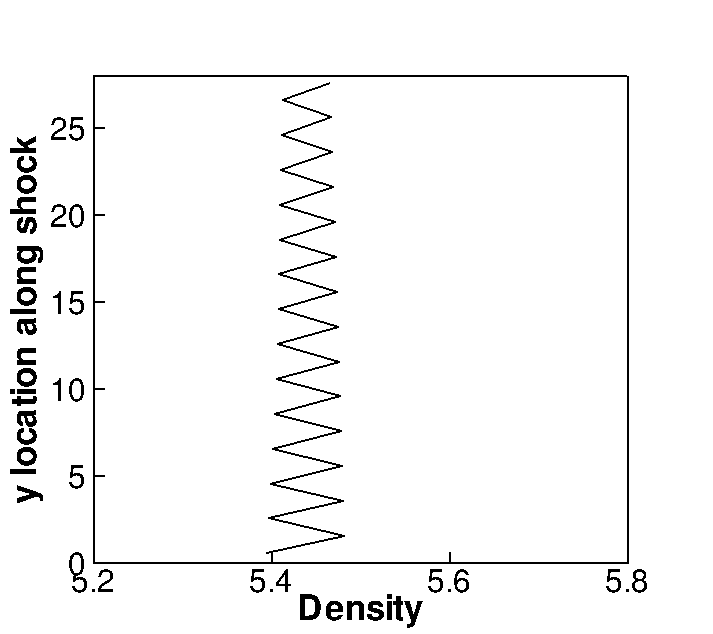}}
	    \subfloat[HLLEM-SWM-P]{\label{fig:hllem_swm_p_inclinedshock_densityalongshock}\includegraphics[scale=0.3]{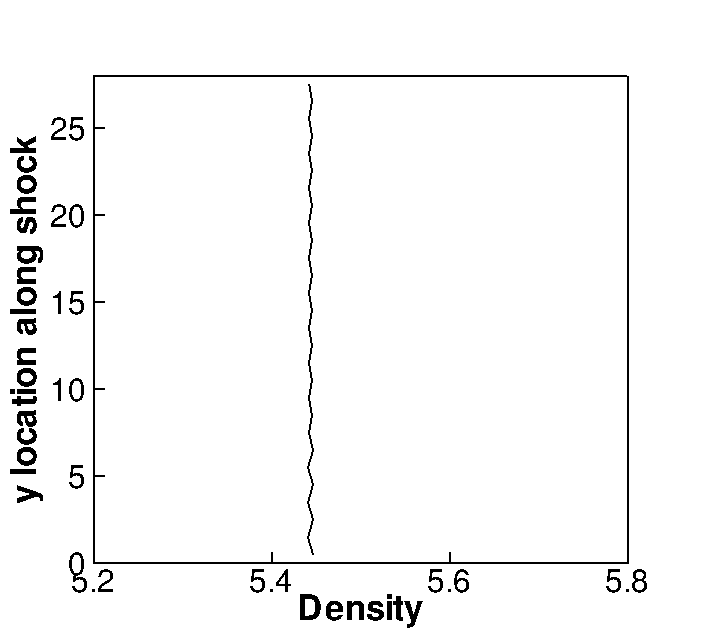}}\\
	    \subfloat[HLLEM-SWM-E]{\label{fig:hllem_swm_e_inclinedshock_densityalongshock}\includegraphics[scale=0.3]{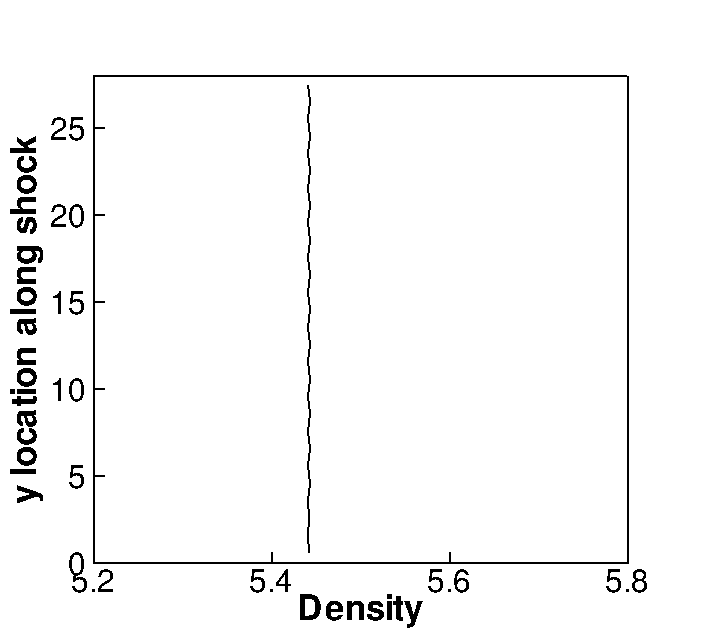}}
	    \subfloat[HLLEM-ADC]{\label{fig:hllem_adc_inclinedshock_densityalongshock}\includegraphics[scale=0.3]{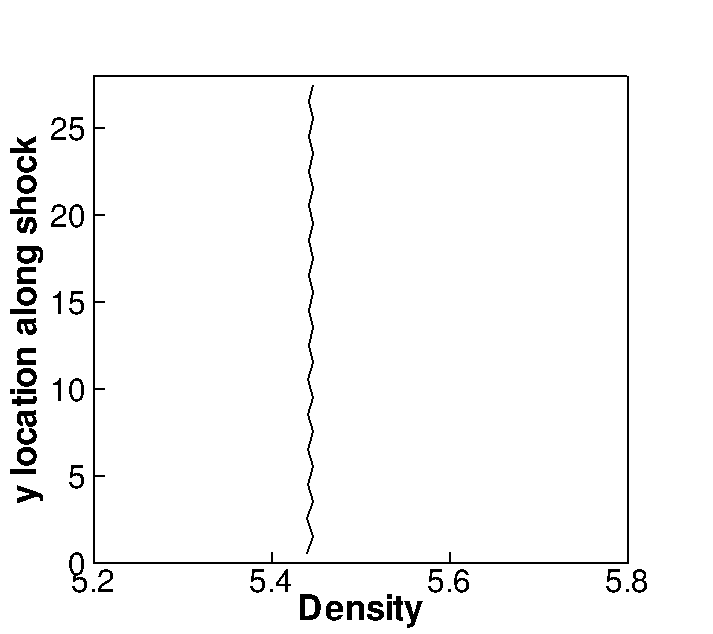}}\\
	    \subfloat[HLLEMS]{\label{fig:hllems_inclinedshock_densityalongshock}\includegraphics[scale=0.3]{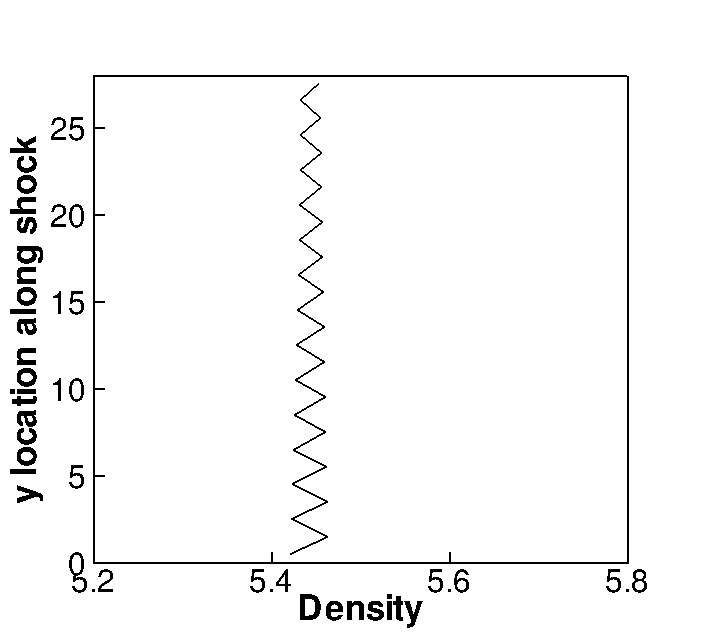}}
	   \caption{Density contours for M=7 stationary inclined shock problem.}
	   \label{fig:inclinedshock}
	    \end{figure}

Fig.(\ref{fig:hllem_inclinedshock}) shows the presence of instability in the HLLEM scheme. Fig.(\ref{fig:hllem_inclinedshock_densityalongshock}) shows the density variation 
along the shock front extracted just behind the shock. The density variation has a typical saw tooth profile that is characteristic of a shock unstable solution. 
Figs.(\ref{fig:hllem_swm_p_inclinedshock},\ref{fig:hllem_swm_e_inclinedshock}) show that both the HLLEM-SWM variants are able to fix this issue providing a clean 
shock profile. The respective density variations in Fig.(\ref{fig:hllem_swm_p_inclinedshock_densityalongshock},\ref{fig:hllem_swm_e_inclinedshock_densityalongshock}) confirm this observation. 
It is interesting to see from Figs.(\ref{fig:hllem_adc_inclinedshock},\ref{fig:hllems_inclinedshock}) that amongst the HLLEM fixes that aims to control the antidiffusion 
terms to achieve shock stability, only the HLLEM-ADC variant is able to completely save the HLLEM scheme from instability. The density variation behind shock shown in 
Fig.(\ref{fig:hllems_inclinedshock_densityalongshock}) corresponding to the HLLEMS scheme indicates that controlling 
only the shear component may not suffice in curing instability. Hence, 
for a complete removal of instability in case of HLLEM scheme, total control of the antidissipation vector $\mathbf{A}$ has to be achieved as observed from the analysis is 
Section.(\ref{sec:orderofmagnitudeanalysis_hlle_hllem}).

\subsection{Laminar flow over flat plate (Viscous)}
\label{sec:flatplate_hllemfixed}

\begin{figure}[H]
	    \centering
	     \subfloat[]{\label{fig:laminarflatplate}\includegraphics[scale=0.4]{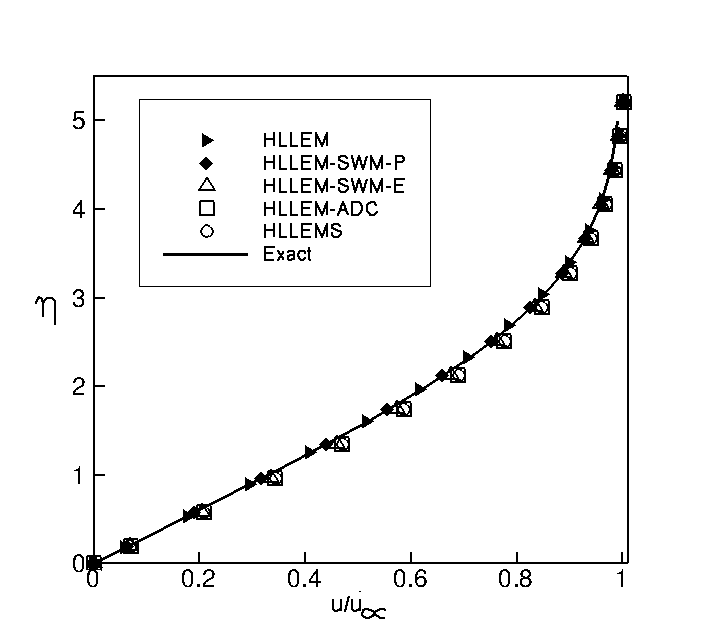}}\\
	    \subfloat[]{\label{fig:laminarflatplate_zoomed}\includegraphics[scale=0.4]{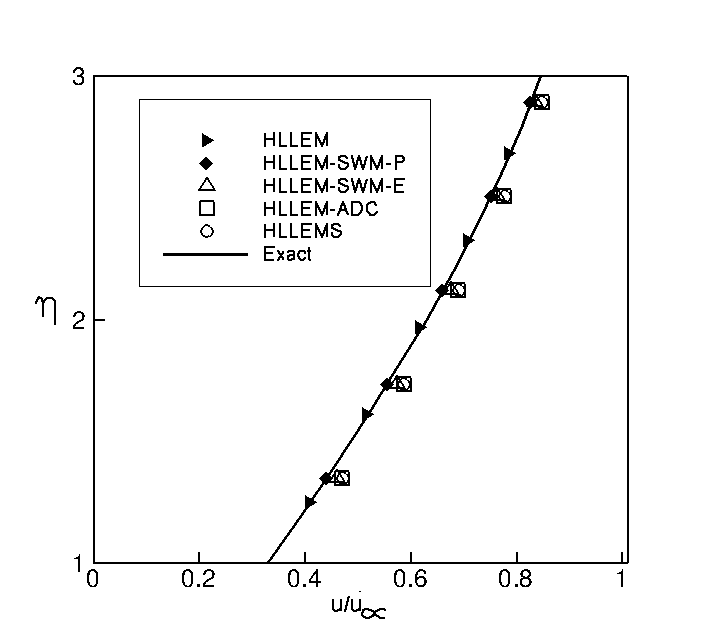}}
	   \caption{(a)Normalized longitudinal velocity variation for subsonic laminar flow over a flat plate (b) Zoomed-in plot showing variation of Blasius parameter $\eta$ between $1$ and $3$. }
	   \label{fig:laminarflatplate_overall}
	    \end{figure}

Since all the cures discussed in Sec.(\ref{sec:formulations}) are based on introducing some or the other form of dissipation into HLLEM scheme, it is imperative to check 
how these extraneous dissipation affects resolution of shear layers. To this end, the classic test case of a laminar flow over a flat plate is used. 
The problem is set up as follows: A flow with $M = 0.1$, pressure of 41368.5 Pa, temperature of 388.88 K, constant dynamic 
viscosity $\mu$ of 2.23$e^{-5} \frac{Ns}{m^2}$ , gas constant of 287.0 $\frac{J}{Kg K}$, coefficient of specific heat at constant pressure Cp of
1005.0 $\frac{J}{Kg K}$ and Prandlt number of 0.72 is computed over a flat plate of length L=0.3048 m. The total length of the domain 
is 0.381 m in x direction and 0.1 m in y direction. The domain is divided into 31$\times$33 Cartesian cells. While uniform meshing is 
done in the x direction, a non-uniform grid spacing is preferred in the y direction with atleast 15 cells within the boundary layer.
Viscous fluxes were discretized using the Coirier diamond path method discussed in \cite{coirier1994,mandal2011}. CFL was taken to be 0.7. 
The flow was considered to have achieved steady state when the horizontal velocity residuals dropped to the order of $1E^{-7}$. 
The normalized longitudinal velocity profiles $(\frac{u}{u_{\infty}})$ are plotted against the Blasius parameter $\eta=y\sqrt{u_{\infty}/\mu L}$ in 
Fig.(\ref{fig:laminarflatplate_overall}).  
It is clearly seen from Fig.(\ref{fig:laminarflatplate}) that the HLLEM-SWM-P is able to exactly resolve the boundary layer as well as the HLLEM scheme. The HLLEM-SWM-E scheme is 
also quite accurate although it is incapable of exact resolution. The HLLEM-ADC and the HLLEMS schemes are both dissipative compared to these HLLEM-SWM class of schemes. This is
quite evident in Fig.(\ref{fig:laminarflatplate_zoomed}) where a zoomed-in plot is shown with $1\leq\eta\leq3$. It can be inferred that slightly away from the wall, the pressure
sensor $\omega$ employed in these schemes are active on the transverse interfaces due to variations in pressure along the plate leading to unwanted reduction in quantity of
antidiffuisve terms. However close to the wall, all cures are found to be capable of satisfactorily resolving the velocity gradient.
\section{Conclusions}
\label{sec:conclusions}
In this paper we presented some strategies to construct various versions of shock stable HLLEM scheme. We began by analyzing the numerical dissipation 
characteristics of the HLLE and the HLLEM schemes in the vicinity of a normal shock using a linear scale analysis technique. 
It was observed that in the transverse direction of a normal shock subjected to numerical perturbations, 
the antidiffusive terms in the mass and interface-normal momentum flux discretizations of the HLLEM scheme gets activated which in turn causes a weakening 
of its inherent HLL-type dissipation. The overall reduced dissipation of the
HLLEM scheme is found to be incapable of attenuating perturbations in $\rho$ and $u$ quantities which leads to unphysical variation of conserved quantity $\rho u$ along the length of the 
shock profile and eventually to a shock unstable
solution. 
Based on this understanding, two different strategies were applied in the vicinity of a shock. The first method termed HLLEM-SWM 
aimed to increase 
the magnitude of the dissipation of the inherent HLL component in the mass and interface-normal momentum flux discretizations 
by careful manipulation of certain non-linear wave speed estimates appearing in the HLL-type diffusion vector. The quantity 
of additional dissipation to be infused was obtained through local solution dependent multidimensional shock sensors based on eigenvalues or on pressure ratios.  
However this cure is more effective when applied on all interfaces near a shock wave; ie. along and across a shock front. 
The second method termed HLLEM-ADC, instead, aimed to directly control these critical antidiffusive terms using a pressure based shock sensor and need to be applied only
on the interfaces along the shock front. This makes it the cheapest method amongst the cures discussed here.    
A linear perturbation analysis of these cures revealed the differences in the damping mechanisms they employ to ensure that the perturbations in $\rho$ and $u$ quantities 
are attenuated.
A von-Neumann type stability bounds on the CFL number that is applicable for their effectiveness was derived.
It was also noticed that the HLLEMS scheme, which treats only 
the antidiffusive terms corresponding to shear waves, is incapable of damping perturbations in density variable and hence is instability prone.
A suite of stringent numerical test cases were used to demonstrate the efficacy and robustness of the methods discussed in this paper. It was found that the proposed 
schemes perform
better than HLLEMS scheme for certain numerical test examples like stability of an inclined stationary shock and hypersonic flow over a bluntbody.

\bibliographystyle{ieeetr}
\vspace{-2mm}
\bibliography{References}

\end{document}